\documentclass[twocolumn, prd, aps, nofootinbib, superscriptaddress]{revtex4-1}
\usepackage[utf8]{inputenc}
\usepackage[english]{babel}
\usepackage{amsmath}
\usepackage{mathrsfs}
\usepackage{amssymb}
\usepackage{graphicx}
\usepackage{physics}
\usepackage{hyperref}
\usepackage{xcolor}
\usepackage{comment}
\usepackage{aas_macros}
\begin{document}

\title{Black hole spectroscopy horizons for current and future gravitational wave detectors}

\author{Iara Ota}
\email[]{iara.ota@ufabc.edu.br}
\affiliation{Centro de Ci\^encias Naturais e Humanas, UFABC, Santo Andr\'e, SP  09210-170, Brazil}
\author{Cecilia Chirenti}
\email[]{chirenti@umd.edu}
\affiliation{Department of Astronomy, University of Maryland, College Park, Maryland 20742, USA}
\affiliation{Astroparticle Physics Laboratory NASA/GSFC, Greenbelt, Maryland 20771, USA}
\affiliation{Center for Research and Exploration in Space Science and Technology, NASA/GSFC, Greenbelt, Maryland 20771, USA}
\affiliation{Center for Mathematics, Computation and Cognition, UFABC, Santo Andr{\'e} - SP, 09210-580, Brazil}

\begin{abstract}
Black hole spectroscopy is the proposal to observe multiple quasinormal modes in the ringdown of a binary black hole merger. In addition to the fundamental quadrupolar mode, overtones and higher harmonics may be present and detectable in the gravitational wave signal, allowing for tests of the no-hair theorem. We analyze in detail the strengths and weaknesses of the standard Rayleigh criterion supplied with a Fisher matrix error estimation, and we find that the criterion is useful, but too restrictive. Therefore we motivate the use of a conservative high Bayes factor threshold to obtain the black hole spectroscopy horizons of current and future detectors, i.e., the distance (averaged in sky location and binary inclination) up to which one or more additional modes can be detected and confidently distinguished from each other. We set up all of our searches for additional modes starting at $t = 10(M_1+M_2)$ after the peak amplitude in simulated signals of circular nonspinning binaries. An agnostic multimode analysis allows us to rank the subdominant modes: for nearly equal mass binaries we find $(\ell, m, n) = (2,2,1)$ and $(3,3,0)$ and, for very asymmetric binaries, $(3,3,0)$ and $(4,4,0)$, for the secondary and tertiary modes, respectively. At the current estimated rates for heavy stellar mass binary black hole mergers, with primary masses between 45 and 100 solar masses, we expect an event rate of mergers within our conservative estimate for the $(2,2,1)$ spectroscopy horizon of $0.03 - 0.10\ {\rm yr}^{-1}$ for LIGO at design sensitivity and $(0.6 - 2.4) \times 10^3\ {\rm yr}^{-1}$ for the future third generation ground-based detector Cosmic Explorer.
\end{abstract}

\maketitle

\section{Introduction}
\label{sec:Intro}
The increasing number of detected binary black hole (BH) mergers has turned gravitational wave (GW) astronomy into a multi-faceted new field, with important ramifications for astrophysics and cosmology, among other disciplines. In the future, low-frequency ($\lesssim 1 {\rm Hz}$) and high-frequency ($\gtrsim 1 {\rm kHz}$) GW astronomers may become as specialized as their current radio and X-ray counterparts. However, some physical features will be the same over the entire frequency spectrum and remain relevant for all future GW astronomers. According to Einstein's theory of General Relativity (GR), BHs are \emph{scale-invariant} (although corrections to GR need not be \cite{2013LRR....16....9Y}): the properties of  Kerr spacetime are the same for the lightest stellar mass BH \cite{2020ApJ...896L..44A} up to the heaviest behemoth imaginable \cite{2019ApJ...887..195M}.

These properties are encoded in the BH spectrum of discrete quasinormal mode (QNM) complex frequencies $\omega_{\ell m n} = \omega^r_{\ell m n} + i \omega^i_{\ell m n}$ \cite{Kokkotas:1999bd,Nollert_1999,Berti:2009kk}, where the numbers $(\ell, m)$ identify the angular symmetry of each mode and the number $n$ is the tone. The no-hair theorem guarantees that astrophysical BHs are described exclusively by their mass $M$ and spin parameter $a$; there is no additional fine print in GR. As a consequence, the detection of a single QNM is sufficient  to provide two equations
\begin{equation}
M = M(\omega^r_{\ell m n},\omega^i_{\ell m n}) \quad {\rm and} \quad a = a(\omega^r_{\ell m n},\omega^i_{\ell m n})
\label{eq:nohair}
\end{equation}
to completely determine the BH spacetime. At least in the classical theory, BHs in a vacuum hold no secrets. Such measurements are already possible with the LIGO/Virgo observations \cite{LIGOScientific:2018mvr,LIGOScientific:2020niy}. We can highlight the detections of GW150914 \cite{LIGOScientific.116.061102} (the first and still strongest binary BH event to date) and GW190521 \cite{Abbott:2020tfl} (which featured the heaviest merging BHs observed so far). The fundamental quadrupolar QNM, i.e. the mode labeled with $(\ell,m,n) = (2,2,0)$, was measured in the post-merger ringdown of both events. The properties of the merger remnant obtained in this way can be compared with the results from the analysis of the inspiral part of the GW signal, providing a consistency check of GR \cite{TheLIGOScientific:2016src,Abbott:2020mjq,Abbott:2020jks}.

But the GW community is eager to probe deeper into the nature of BHs. The confident detection of a secondary, or subdominant, QNM would provide additional information to allow an independent test of the no-hair theorem through BH spectroscopy \cite{Dreyer:2003bv,Berti:2005ys,Cardoso:2016ryw,Berti:2018vdi,Brito:2018rfr}.
Higher harmonic modes with $(\ell,m) \ne (2,2)$ become more relevant for more unequal mass binaries, with large mass ratio $q = M_1/M_2 \geq 1$. Moreover, overtone modes ($n>0$) decay much faster than fundamental modes ($n=0$).  Therefore, most studies focused on performing spectroscopy add only fundamental higher harmonic modes, i.e., $(\ell, m, 0) \ne (2,2,0)$ \cite{Berti:2007fi, Kamaretsos:2011um,Thrane:2017lqn,Cotesta:2018fcv,Baibhav:2018rfk,Carullo:2019flw,Maselli:2019mjd,Uchikata:2020wsp,Capano:2020dix,Shaik:2019dym,Capano:2021etf}, even though the importance of overtones has been known for years~\cite{Leaver:1986gd, Stark:55.891} and a few more recent works already indicated the relevance of the first overtone of the quadrupolar mode, labeled (2,2,1)  \cite{Buonanno:2006ui,London:2014cma,Baibhav:2017jhs}.

The recent studies of the importance of higher overtones in the ringdown modeling \cite{Baibhav:2017jhs, Giesler:2019uxc} and some suggestive evidence for the presence of the $(2,2,1)$ mode in GW150914 \cite{Isi:2019aib} increased interest in the contribution of overtones \cite{Bhagwat:2019dtm,Ota:2019bzl,Forteza:2020hbw,Mourier:2020mwa,Okounkova:2020vwu,Bustillo:2020tjf,Finch:2021iip,Isi:2020tac,2021arXiv210705609I,Forteza:2021wfq}.
More recently, some works extended the ringdown modeling analysis to include mirror modes~\cite{Forteza:2020hbw,Dhani:2020nik,Finch:2021iip,Dhani:2021vac}.

It is natural to ask when we will be confidently able to do BH spectroscopy. In particular, will high-precision BH spectroscopy be attainable with LIGO? Motivated by the ringdown horizons computed by~\cite{Baibhav:2018rfk}, we present here the first detailed study of the \emph{BH spectroscopy horizon}, which measures how far an event can be, as a function of the remnant mass, binary mass ratio and the detector being considered, for two or more QNMs to be detectable in the GW ringdown.\footnote{We define the BH spectroscopy horizon with the angular average of sky location and binary inclination (see Section~\ref{sec:Rayleigh}). In the LIGO literature, a ``horizon" is usually defined as the maximum distance obtained with optimal values of the angular parameters, while a ``range" refers to the angular average. Therefore our horizon definition is equivalent to the range definition used by LIGO.}
It is important to stress that we are not performing tests of GR in our work. Rather, we are assessing the detectability of a secondary mode, assuming that GR is the correct theory of gravity.
All of our analyses start at a conservative time $t = 10(M_1 + M_2)$ after the signal peak.

We probe this topic with two different approaches. First, we use a Rayleigh criterion \cite{Berti:2005ys,Berti:2007zu} coupled with a Fisher matrix analysis \cite{Finn:1992wt} to estimate the spectroscopy horizon for LIGO at design sensitivity \cite{Design-sensitivity}, future space detector LISA \cite{amaroseoane2017laser} and proposed third generation ground-based detectors Einstein Telescope (ET) \cite{2010CQGra..27s4002P} and Cosmic Explorer (CE) \cite{2017CQGra..34d4001A} (see Section \ref{sec:Rayleigh}). Second, we use a fully Bayesian analysis to verify our results for LIGO and CE, using a well-motivated and conservative Bayes factor threshold (Section \ref{sec:Bayes}). We confirm that the Rayleigh criterion is too restrictive (see Section \ref{sec:discussion}), and find promising rates of detection with third generation detectors (and LISA). The LIGO rates are small, but not prohibitively so. We present our conclusion in Section \ref{sec:conclusions} and also explore other prescriptions for mode resolvability in the Appendix \ref{sec:appendix-other-criteria}.

\section{BH spectroscopy horizons from the Rayleigh criterion}
\label{sec:Rayleigh}

Our goal is to determine whether two modes are detectable in the ringdown of a binary BH coalescence at a given luminosity distance $D_L$, with binary component masses $M_1 > M_2$ resulting in a merger remnant black hole of mass $M$ and spin parameter $a$.

The ringdown gravitational waveform can be written as a sum of QNMs
\begin{align}
h_{+} + ih_{\times}
&=\frac{M}{D_L}\sum_{\ell mn}A_{\ell mn}e^{i[\omega_{\ell mn}(t-t_0)-\phi_{\ell mn}]}\nonumber\\
&\times {}_{-2}S_{\ell m}(a\omega_{\ell mn},\iota,\beta),
\label{eq:qnm}
\end{align}
for $t > t_0$, a time close but not necessarily equal to the merger time of the binary. Here $h_+$ and $h_{\times}$ are the two polarizations of the wave, $A_{\ell mn}$ and $\phi_{\ell mn}$ are the initial amplitude and phase of each mode, respectively, $\omega_{\ell mn} = \omega^r_{\ell mn} + i \omega^i_{\ell mn}$ are the complex frequencies of the modes, ${}_{-2}S_{\ell m}(a\omega_{\ell mn},\iota,\beta)$ are the spin-weighted spheroidal harmonics with spin $s = -2$, and $(\iota,\beta)$ are the inclination and azimuth angles of the binary relative to the detector. The frequencies of oscillation and damping times of the modes are given by
\begin{equation}
f_{\ell mn} = \frac{\omega^r_{\ell mn}}{2\pi}\quad {\rm and} \quad \tau_{\ell mn} = \frac{1}{\omega^i_{\ell mn}}.
\label{eq:f_tau}
\end{equation}

The physical quantities shown in eqs. (\ref{eq:qnm}) and (\ref{eq:f_tau}) are measured in the source frame. The corresponding values measured in the detector frame are rescaled by the cosmological redshift $z$: $M^{\rm d} = M(1+z)$, $f^{\rm d}_{\ell mn} = f_{\ell mn}/(1+z)$ and $\tau^{\rm d}_{\ell mn} = \tau_{\ell mn}(1+z)$, where the superscript ``d'' indicates that the quantities are measured in the detector frame. Whenever required, we use values of the cosmological parameters given in \cite{2020A&A...641A...6P}.

The waveform measured by the detector is given by \cite{Moore:2014lga}
\begin{equation}
	h(t) = F_{+}(\theta,\phi,\psi)h_{+}(t) + F_{\times}(\theta,\phi,\psi)h_{\times}(t),
	\label{eq:antenna}
\end{equation}
where $h_{+,\times}(t)$ are the wave polarizations defined in equation~\eqref{eq:qnm} with quantities measured in the detector frame, $F_{+,\times}(\theta,\phi,\psi)$ are the detector's antenna patterns, $(\theta,\phi)$ indicate the source sky location and $\psi$ is the polarization angle. We use the full-sky angular  average of the spin-weighted spheroidal harmonics $\langle | {}_{-2}S_{\ell m}| \rangle = 1/\sqrt{4\pi}$ and of the antenna patterns $\langle (F^2_{+,\times})^{\frac{1}{2}}\rangle$, which are equal to $1/\sqrt{5}$ for L shaped detectors (LIGO and CE) \cite{Moore:2014lga} and $3/(2\sqrt{5})$ for ET~\cite{Regimbau:2012ir} and does not have a closed analytical form for LISA~\cite{Robson:2018dyw}.

\begin{figure}[htb!]
	\centering
	\includegraphics[width = 1.0\linewidth]{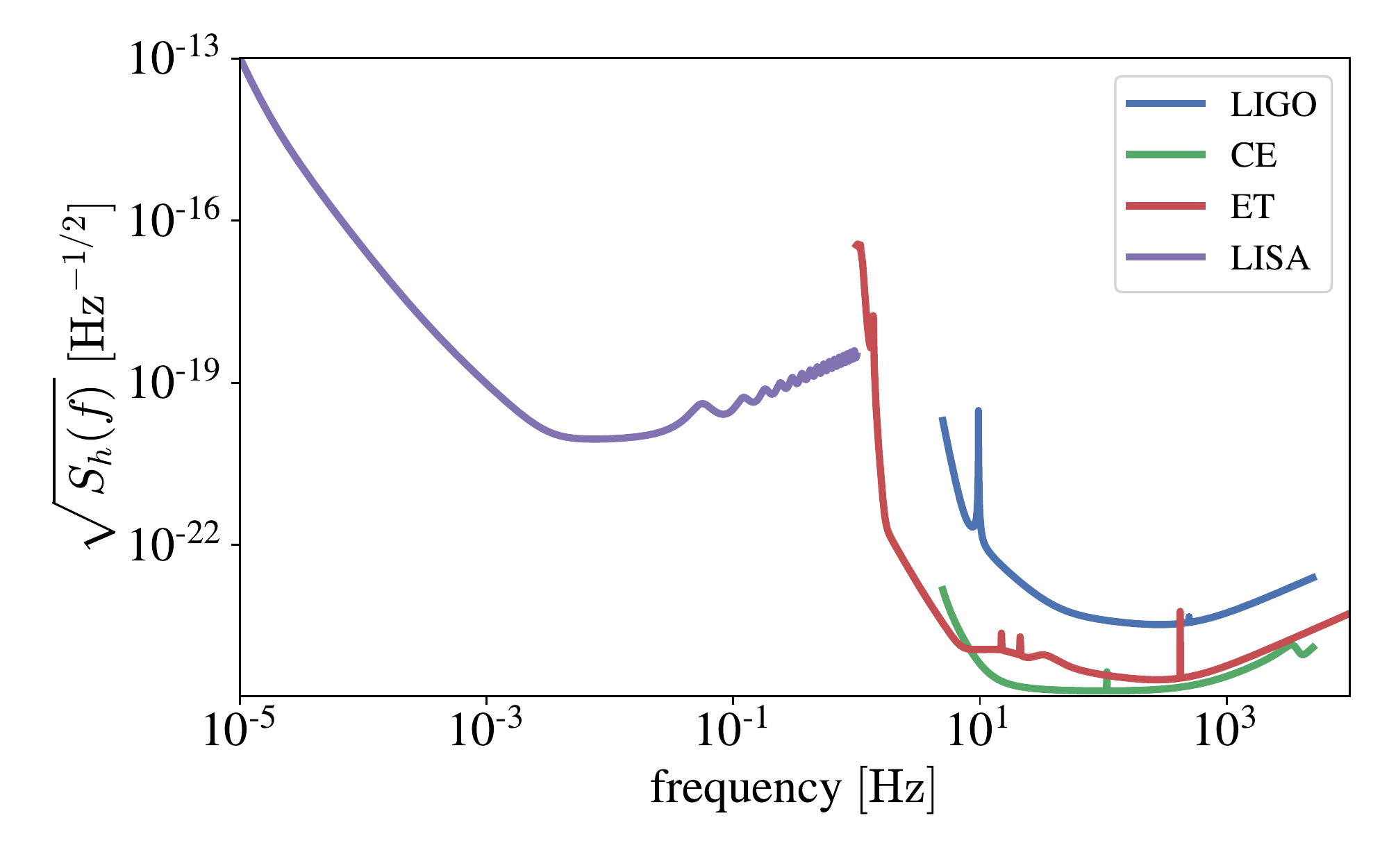}
	\caption{Noise spectral density for current and future GW detectors used in our study: LIGO (at design sensitivity)~\cite{Design-sensitivity}, proposed third generation ground-based detectors Cosmic Explorer (CE)~\cite{CE_noise_curve} and Einstein Telescope (ET)~\cite{ET_noise_curve, 2010CQGra..27s4002P} and space-based detector LISA \cite{amaroseoane2017laser} (planned launch in mid 2030's).}
	\label{fig:detector_noise}
\end{figure}

We also need to define the single detector signal-to-noise ratio (SNR), given by
\begin{equation}
\rho^2 = 4\int_0^{\infty}\frac{|\tilde{h}(f)|^2}{S_n(f)}df,
\end{equation}
where the symbol (\textasciitilde) denotes the Fourier transform of the function and $S_n(f)$ is the noise spectral density of the detector. In Figure \ref{fig:detector_noise} we show $S_n(f)$ for LIGO (at design sensitivity) and future detectors CE, ET and LISA. 

For our spectroscopic analysis, we require that each mode $(\ell,m,n)$ has a single detector SNR $\rho_{\ell mn} > 8$ and appraise their detectability using a Rayleigh criterion (Section \ref{sec:criterion}) with errors estimated by a Fisher matrix analysis (Section \ref{sec:Fisher}). The spectroscopy horizons are presented in Section \ref{sec:rayleigh-modes-horizions}. 

\subsection{Rayleigh criterion and caveats}
\label{sec:criterion}

The original Rayleigh criterion serves the purpose of establishing whether two close stars are resolvable in an optical telescope. In direct analogy, Berti et al. \cite{Berti:2005ys} proposed an adaptation of the criterion to determine whether two quasinormal modes $(\ell,m,n)$ and $(\ell',m',n')$ can be resolved in a GW detection. The required conditions are:
\begin{subequations}
\begin{align}
\label{eq:rayleigh_f}
\Delta f_{\ell mn,\ell'm'n'}  > {\rm max}(\sigma_{f_{\ell mn}},\sigma_{f_{\ell' m'n'}}),\\
\Delta \tau_{\ell mn,\ell'm'n'} > {\rm max}(\sigma_{\tau_{\ell mn}},\sigma_{\tau_{\ell' m'n'}}),
\label{eq:rayleigh_tau}
\end{align}
\label{eq:rayleigh_both}
\end{subequations}
where we use the definitions
\begin{subequations}
\begin{align}
\Delta f_{\ell mn,\ell'm'n'} \equiv |f_{\ell mn} - f_{\ell'm'n'}|,\\
\Delta \tau_{\ell mn,\ell'm'n'} \equiv |\tau_{\ell mn} - \tau_{\ell'm'n'}| ,
\end{align}
\end{subequations}
and we assume that each frequency and damping time are reported from the observations as $f_{\ell mn} \pm \sigma_{f_{\ell mn}}$ and  $\tau_{\ell mn} \pm \sigma_{\tau_{\ell mn}}$, respectively. This criterion has been used in different works to estimate the resolvability of two QNMs \cite{Berti:2007zu,Bhagwat:2019dtm,Ota:2019bzl,Forteza:2020hbw}.

We choose $(\ell,m,n) = (2,2,0)$ (the fundamental mode), and let $(\ell',m',n')$ be a subdominant mode, that is, either the first overtone $(2,2,1)$ or one of the fundamental higher harmonics. If \emph{both} conditions (\ref{eq:rayleigh_both}) are satisfied, then we can say that both pairs of frequencies and damping times are independently determined. This allows a test of the no-hair theorem using eq. (\ref{eq:nohair}) for each mode.

\begin{figure}[htb!]
	\centering
	\includegraphics[width = 1.0\linewidth]{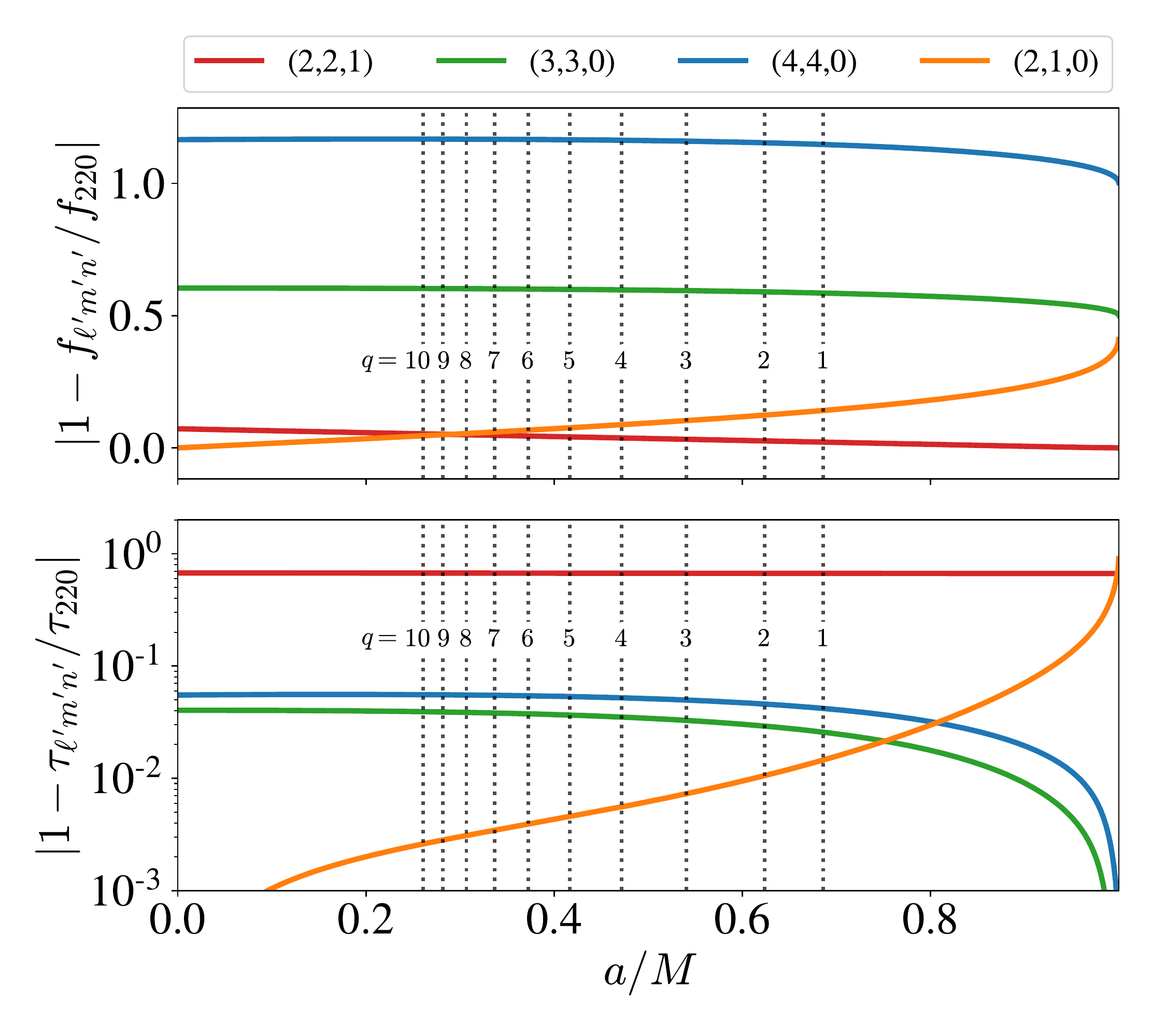}
	\caption{Relative differences of frequencies (top) and damping times (bottom) between the dominant mode $(2,2,0)$ and the most relevant subdominant modes
	as a function of the final black hole dimensionless spin $a/M$, obtained with linear perturbation theory~\cite{Berti:2005ys,Berti-ringdown}. The vertical dotted lines indicate the final dimensionless spin of the remnant of nonspinning circular binaries with mass ratios $q$ raging from $1$ (highest spin) to $10$ (lowest spin). The modes with $\ell = 2$ have frequencies very similar to the $(2,2,0)$ frequency, whereas modes with $\ell \ne 2$ have more distinct (higher) frequencies, which are more easily resolvable with eq. (\ref{eq:rayleigh_f}). The overtone has a lower damping time that is more easily resolvable with eq. (\ref{eq:rayleigh_tau}) than any of the fundamental modes, which have similar damping times to the $(2,2,0)$ mode.}
	\label{fig:QNMs_a_q}
\end{figure}

As we can see from Fig. \ref{fig:QNMs_a_q}, the Rayleigh conditions (\ref{eq:rayleigh_both}) pose different challenges for the overtone (condition (\ref{eq:rayleigh_tau}) is easily satisfied) and the higher harmonics with $\ell \ne 2$ (condition (\ref{eq:rayleigh_f}) is easily satisfied).
There is also a dependence on the mass ratio of the progenitor BH binary: for nonspinning circular binaries with a fixed total mass, the final spin $a$ of the merger remnant will depend only on the binary mass ratio, due to its relation to the orbital angular momentum. Therefore the final spin is higher for equal mass binaries.

For modes with $\ell = m$, both $\Delta f_{\ell mn,\ell'm'n'} $ and $\Delta \tau_{\ell mn,\ell'm'n'}$ increase with $q$, which helps satisfy the Rayleigh criterion~\eqref{eq:rayleigh_both}.
But binaries with higher mass ratio excite the dominant mode with lower amplitude and emit less gravitational energy for a fixed total mass (see Table \ref{table:qnm_pars}), leading to lower SNR and larger measurement errors $\sigma_{f_{\ell mn}}$ and $\sigma_{\tau_{\ell mn}}$.
The combination of these effects results in a \emph{smaller} spectroscopy horizon of the modes with $\ell = 2$ for higher mass ratios. For the higher harmonics with $\ell = m \ne 2$ the spectroscopy horizon is approximally the same for low and high mass ratio, as we show in Section ~\ref{sec:rayleigh-modes-horizions}.

Two caveats must be mentioned about the Rayleigh criterion. First, the standard implementation we present here gives an either-or answer (the two modes are either resolvable or not) and permits no gradation. (See also a discussion of variations of the Rayleigh criterion that are equivalent to different confidence levels in Appendix \ref{sec:appendix-other-criteria}.) Second, the criterion could be inherently \emph{too restrictive} for predicting the detectability of a secondary mode, i.e., it could be a stronger condition than what is actually needed for identifying a secondary mode in the data with a reasonable level of confidence (see Sec. \ref{sec:discussion}).\footnote{Both caveats were also addressed by~\cite{Berti:2007zu}, where a generalized likelihood ratio test (GLRT) was proposed to determine whether a secondary mode is present or not in a signal, finding a SNR higher than the SNR needed to resolve a single Raleigh condition but smaller than the SNR needed to resolve both conditions, see their Appendix B. This is similar to the approach we use in Section~\ref{sec:Bayes} with a Bayes factor threshold.}
Previous work showed that the minimum ringdown SNR needed to resolve both Rayleigh conditions  is very high~\cite{Berti:2005ys, Bhagwat:2019dtm,Ota:2019bzl} (${\rm SNR} \approx 100$).
On the other hand, the minimum SNR for satisfying just one condition of the Rayleigh criterion may not be restrictive enough for a confident detection, and usually needs to be complemented with additional conditions, such as a minimum SNR threshold~\cite{Berti:2005ys, Bhagwat:2016ntk,Forteza:2020hbw}.

\subsection{Fisher matrix analysis}
\label{sec:Fisher}

For a high SNR measurement, the probability distribution of each parameter $\vartheta^a$ of the waveform will be a Gaussian centered at the real value $\vartheta^a_{\rm real}$. Consequently, the estimated value for parameter $\vartheta^a$ will be close to the real value $\vartheta^a_{\rm real}$, that is, $\vartheta^a = \vartheta^a_{\rm real} + \delta \vartheta^a$ and the probability density distribution for $\delta \vartheta^a$ is proportional to the multidimensional Gaussian $\exp[-(1/2)\Gamma_{ab}\vartheta^a\vartheta^b]$. The statistical error in the determination of $\vartheta^a$ is
\begin{equation}
\sigma_{\vartheta^a} = \sqrt{(\Gamma^{-1})^{aa}},
\label{eq:sigma}
\end{equation}
where $\Gamma^{-1}$ is the inverse of the Fisher matrix \cite{Finn:1992wt}, given by
\begin{equation}
\Gamma_{ab} \equiv \left\langle\frac{\partial h}{\partial \vartheta^a}\left|\frac{\partial h}{\partial \vartheta^b}\right.\right\rangle,
\label{eq:fisher-matrix}
\end{equation}
with the noise-weighted inner product defined as
\begin{equation}
\langle h_1|h_2\rangle \equiv 2 \int_0^{\infty} \frac{\tilde{h}_1^*(f)\tilde{h}_2(f) + \tilde{h}_1(f)\tilde{h}_2^*(f)}{S_n(f)}df,
\label{eq:product}
\end{equation}
and the symbol (${}^*$) denotes the complex conjugate of the function.

The errors $\sigma_{f_{\ell mn}}$ and $\sigma_{\tau_{\ell mn}}$ needed for the Rayleigh conditions (\ref{eq:rayleigh_both}) can be estimated in this formalism. We choose as our relevant parameters
the QNM intrinsic parameters
$$\vartheta^a = \{f_{\ell mn},\tau_{\ell mn}, A_{\ell mn}, \phi_{\ell mn}, f_{\ell' m'n'},\tau_{\ell' m'n'}, R, \phi_{\ell' m'n'}\},$$
 where $R = A_{\ell' m'n'}/A_{\ell mn}$ is the amplitude ratio between the modes, which gives us 8 parameters for a 2-mode analysis.
We take the full-sky angular average for the sky localization and binary inclination angle relative to the detector (see discussion following eq. (\ref{eq:antenna})).

Following~\cite{Berti:2005ys}, we use the doubling prescription by Flanagan and Hughes \cite{Flanagan:1997sx} to compute the Fourier transform of $h(t)$. That is, we first obtain $h(t)$ for all $t$ by reflecting the QNM waveform at $t_0$; this procedure avoids the spectral leakage at high frequencies that would be caused by a discontinuity in the waveform.
The partial derivatives in equation~\eqref{eq:fisher-matrix} are obtained analytically and the noise-weighted inner product (\ref{eq:product}) and statistical errors given by (\ref{eq:sigma}) are computed numerically.

We use numerical relativity simulations of binary BH mergers from the Simulating eXtreme Spacetimes project (SXS) \cite{Boyle:2019kee,SXS-catalog} to fit the QNM amplitudes and phases for each waveform analyzed here. The procedure for obtaining these fits is detailed in \cite{Ota:2019bzl}, where we examined their dependence on the binary mass ratio. In short, the fundamental modes $(\ell,m,0)$ are fitted starting at the earliest time when the contribution of overtones and non-linearities can be neglected; the overtones are fitted at the time when the waveform is best described by the fundamental mode and the first overtone. The initial amplitudes and phases are then rescaled to the time $t = t_{\rm peak} + 10(M_1 + M_2)$, where $t_{\rm peak}$ is the time of maximum amplitude of the quadrupolar mode $(\ell,m) = (2,2)$.
In Table~\ref{table:qnm_pars} we show the parameters of the QNMs obtained from simulations SXS:BBH:0593 ($q = 1.5$) and SXS:BBH:1107 ($q=10$) \cite{SXS-catalog}, which we analyze in more detail in the following sections.

\begin{table}[htb!]
\centering
\caption{QNM amplitudes, phases, and frequencies (defined in eq. (\ref{eq:qnm})) for nonspinning circular binaries with mass ratio $q=1.5$ and $q=10$. The amplitudes and phases are reported at time $t = t_{\rm peak} + 10(M_1 + M_2)$. The values for the amplitude and phase of the $(2,2,1)$ mode are obtained with method II as described in \cite{Ota:2019bzl}.}
\begin{ruledtabular}
\begin{tabular}{c c c c c}
$(\ell,m,n)$ & $A_{\ell mn}$ & $\phi_{\ell mn}$ [rad]& $\omega^r_{\ell mn} MG/c^3$ & $\omega^i_{\ell mn} MG/c^3$ \\ \hline\hline
\multicolumn{5}{c}{$q = 1.5$}\\ \hline
$(2,2,0)$ & 0.40 & 0.41 & 0.517 & 0.082 \\
$(2,2,1)$ & 0.28 & 4.59 & 0.505 & 0.248 \\
$(3,3,0)$ & 0.05 & 6.16 & 0.821 & 0.084 \\
$(4,4,0)$ & 0.01 & 5.46 & 1.112 & 0.086 \\
$(2,1,0)$ & 0.03 & 5.26 & 0.448 & 0.083 \\ \hline\hline

\multicolumn{5}{c}{$q = 10$}\\ \hline
$(2,2,0)$ & 0.14 & 3.56 & 0.412 & 0.088 \\
$(2,2,1)$ & 0.07 & 1.17 & 0.390 & 0.269 \\
$(3,3,0)$ & 0.05 & 3.32 & 0.661 & 0.092 \\
$(4,4,0)$ & 0.02 & 3.32 & 0.894 & 0.093 \\
$(2,1,0)$ & 0.05 & 5.57 & 0.394 & 0.088 \\
\end{tabular}
\end{ruledtabular}
\label{table:qnm_pars}
\end{table}

\subsection{BH spectroscopy horizons}
\label{sec:rayleigh-modes-horizions}

\begin{figure*}[htb!]
	\centering
	\includegraphics[width = 1.0\linewidth]{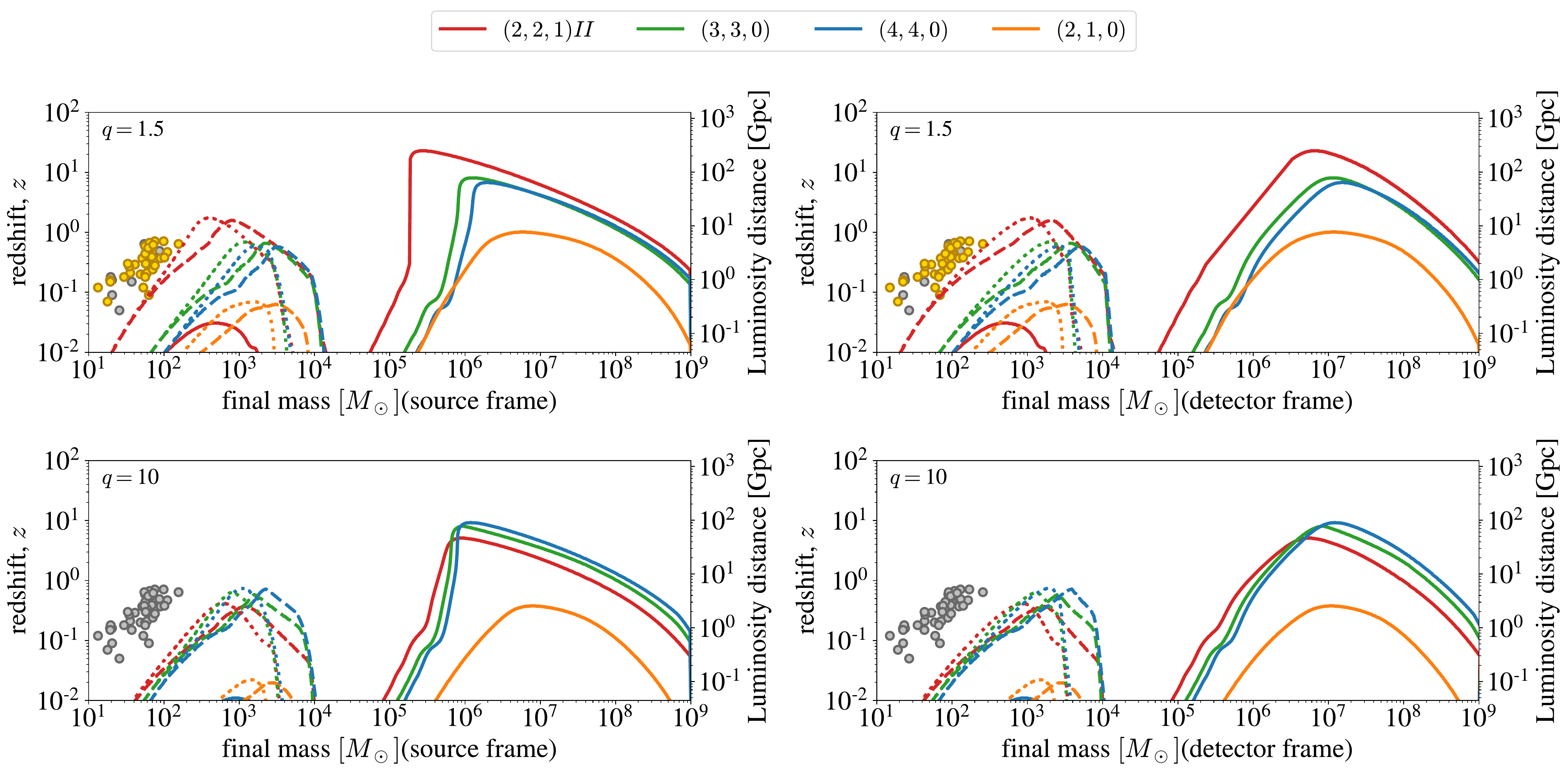}
	\caption{BH Spectroscopy horizons obtained with the Rayleigh criterion (\ref{eq:rayleigh_both}) and Fisher matrix error estimation with 8 parameters. Top (bottom) panels show results for a BH merger remnant from a binary with a low (high) mass ratio $q = 1.5$ ($q = 10$). The left (right) panels show the BH spectroscopy horizons as a function of the BH remnant mass in the source (detector) frame. Solid lines indicate the horizons for LIGO (lower masses) and LISA (higher masses); dashed (dotted) lines indicate the ET (CE) horizons.  Circles show detections from GWTC-1 and GWTC-2; yellow circles are compatible with the mass ratio in each case, whereas gray circles are not. Ringdown modes $(2,2,0) + (\ell,m,n)$ will be detectable and resolvable for events that happen below the corresponding $(2,2,0) + (\ell,m,n)$ spectroscopy horizon curve.}
	\label{fig:horizon_rayleigh}
\end{figure*}

Here we present the BH spectroscopy horizon, $z_{\ell mn}^{\rm spec, R}$, which is the maximum distance, averaged over
sky location and binary inclination, up to which a secondary mode $(\ell,m,n)$ can be detected and resolved in the signal, using the Rayleigh conditions (\ref{eq:rayleigh_both}) supplied with errors given by the Fisher matrix formalism (\ref{eq:sigma}). The spectroscopy horizon provides a figure of merit which allows us to compare results between detectors and across several orders of magnitude in mass range. 

We propose the use of the source distance as a more universal measure for assessing the prospects of BH spectroscopy than the SNR needed for detection, which depends on the detector and the source.\footnote{In~\cite{Bhagwat:2019dtm}, for instance, it was shown that the ringdown SNR of a signal containing the fundamental mode plus the overtone could be lower than the SNR of the fundamental mode alone, depending on the relative phases of the modes.} In short, the SNR is a good measure to determine detectability thresholds, but it is not the best option to use when comparing ringdown signals with different subdominant modes. If all else is kept equal (binary mass ratio, remnant mass, distance, etc) and we inform the properties of the modes (initial amplitudes and phases) with NR simulations, then we find that two signals with different subdominant modes will have different SNRs at the same distance. Therefore, a fixed SNR threshold is not a good equalizer: it requires different source distances for different subdominant modes, leading to different detection rates. This problem is solved by quoting directly the source distance, which we calculate as detailed below to find the BH spectroscopy horizon.

\begin{figure}[htb!]
	\centering
	\includegraphics[width = 1.0\linewidth]{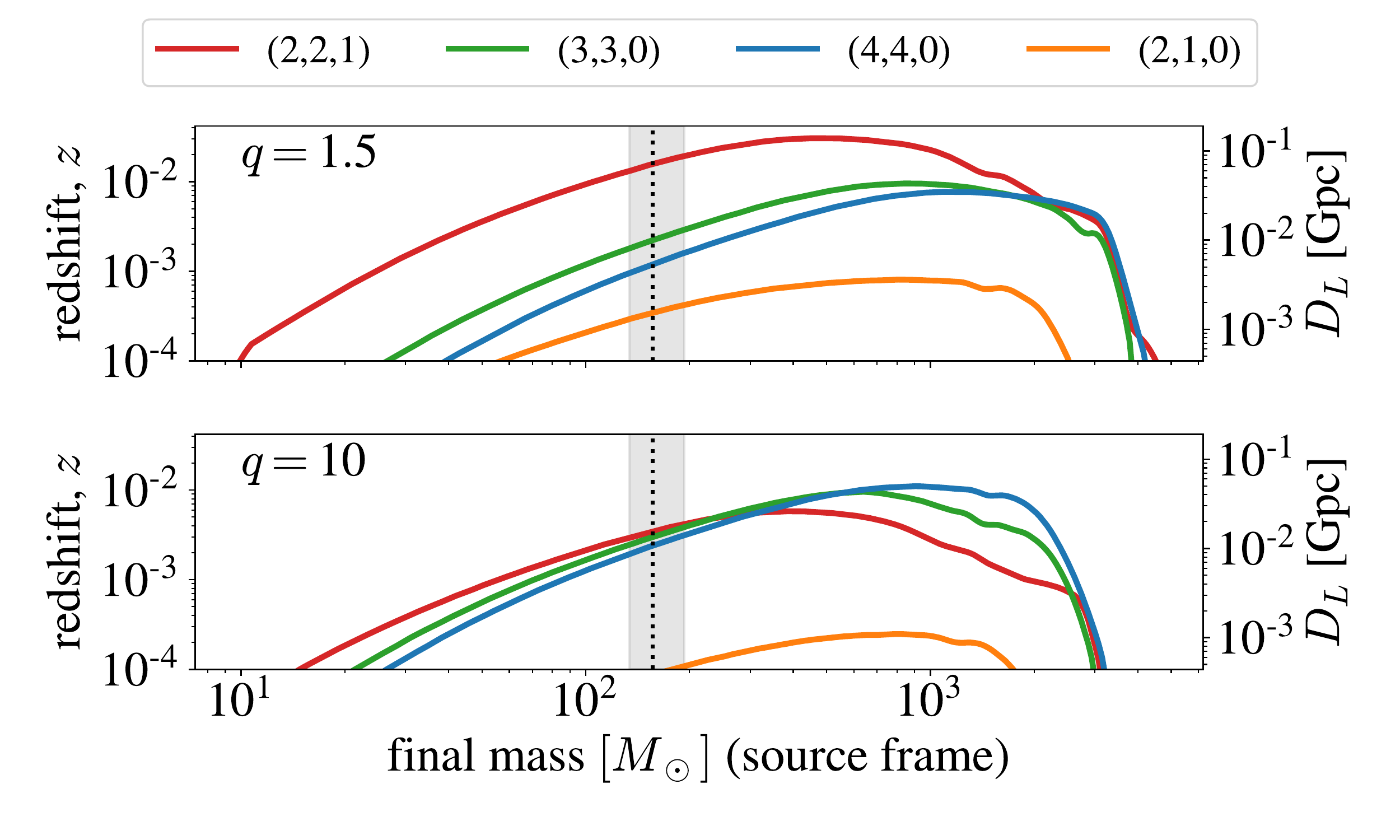}
	\caption{Same as Figure \ref{fig:horizon_rayleigh}, but for the LIGO spectroscopy horizons (at lower distances). The gray band corresponds to the estimated remnant mass of GW190521. The largest corresponding spectroscopy horizon, however, is approximately 10 times closer than the estimated distance to the event.}
	\label{fig:horizon_rayleigh_LIGO}
\end{figure}

Figure~\ref{fig:horizon_rayleigh} shows the spectroscopy horizons for distinguishing the dominant mode $(2,2,0)$ from the most relevant subdominant modes
as a function of the remnant BH final mass for nonspinning circular binaries with low and high mass ratios. In the area inside the curves, each of the two modes has SNR greater than 8 and they satisfy the Rayleigh criterion \eqref{eq:rayleigh_both}. The circles show the confirmed binary BH events to date~\cite{LIGOScientific:2018mvr,LIGOScientific:2020niy}.
In figure~\ref{fig:horizon_rayleigh_LIGO} we have the spectroscopy horizons just for LIGO, which were mostly not visible in Figure~\ref{fig:horizon_rayleigh}. The LIGO horizons have similar trends as the proposed third generation gravitational wave detectors ET and CE.

Some general features can be seen for all of the detectors. The spectroscopy horizon is small for low masses due to the small amplitude of the QNMs in these systems, as the amplitude is proportional to the BH mass. Systems with larger mass have larger amplitudes, however, as the frequencies are proportional to the inverse of the mass, the SNR decreases as the mode frequencies approach the detector's low frequency sensitivity limit. The SNR of the subdominant modes is larger than approximately 30 in all cases, therefore automatically satisfying our detectability criterion (SNR $>$ 8). Although we do not impose a separate ``measurability'' criterion, see \cite{Forteza:2020hbw}, we verify that $\sigma_{f_{\ell mn}}/f_{\ell mn} \lesssim 0.1$ in all cases. The damping time is typically much more uncertain, and $\sigma_{\tau_{\ell mn}}/\tau_{\ell mn} \lesssim 0.4$.

For the ground-based detectors, the spectroscopy horizon of the overtone (2,2,1) is up to $\sim$ 10-80  times larger than the spectroscopy horizon of the higher harmonics, for lower masses and low mass ratio. This result is compatible with the minimum SNR analysis of an event similar to GW150914 in~\cite{Ota:2019bzl}. For low mass ratio and higher masses, the horizons become comparable (except for the (2,1,0), which is still lower). This is again the case for the high mass ratio case, although we can see that the (2,2,1) spectroscopy horizon is slightly larger for lower masses, while the  spectroscopy horizons of the higher harmonics $(4,4,0)$ and $(3,3,0)$ are larger for larger masses. The larger horizons for higher harmonics with $\ell \neq 2$ in the large mass range are expected, as these modes have larger frequencies than the modes with $\ell=2$ (see Table~\ref{table:qnm_pars}).

It is important to notice that the most massive event detected so far is GW190521~\cite{Abbott:2020tfl}, with final mass $M = 156.3^{+36.8}_{-22.4} M_\odot$~\cite{LIGOScientific:2020niy}. This is also the most massive stellar-mass black hole detected to date. As the existence of even more massive stellar-mass black holes is still uncertain, the Rayleigh criterion analysis with Fisher matrix errors shown in  Figures~\ref{fig:horizon_rayleigh} and~\ref{fig:horizon_rayleigh_LIGO} indicates that, for sources compatible with the current detections and ground-based detectors, it will always be easier to distinguish the overtone from the dominant mode than any of the higher harmonics, even for higher mass ratios.
This result is strongly influenced by the requirement that both Rayleigh conditions should be satisfied, which poses a considerable penalty on the higher harmonics, due to the condition on the damping times.
(See Section~\ref{sec:bayes-two-mode}, for the results of our complementary Bayesian analysis.)
For masses larger than the largest stellar-mass black hole detected to date, the higher harmonics $(3,3,0)$ and $(4,4,0)$ have larger spectroscopy horizons than the overtone.

For LISA the situation is still uncertain, as the properties of the population of supermassive binaries are largely unknown. Therefore we cannot safely estimate which subdominant mode will be more easily detected (although the overtone has a larger spectroscopy horizon in more cases). But the LISA spectroscopy horizons are very large and it is probable that more than one extra mode will be detected in a supermassive binary BH merger. There are two noticeable differences between the general trends of the spectroscopy horizons for LISA when compared with the ground-based detectors. For low mass ratios, the $(2,2,1)$ has the largest spectroscopy horizon for all masses, as a result of the expected flatter behavior of the LISA noise curve for high frequencies (see Figure \ref{fig:detector_noise}). Additionally, we can see the spectroscopy horizons grow almost vertically with the distance for lower masses in the source frame.
This results from the interplay between the cosmological redshift and the detector's colored noise: systems with lower masses have higher frequencies, which are redshifted back to the detectors' most sensitive band. This does not happen in the detector frame because there the masses are also rescaled with the cosmological redshift.

Variations of the Rayleigh criteria we have used in this section have been proposed in other works. For example, ~\cite{Forteza:2020hbw} proposed the use of the quality factor $Q_{\ell m n} = \pi f_{\ell m n} \tau_{\ell m n}$ instead of the damping time $\tau_{\ell m n}$ in the Rayleigh criterion. Requiring only one of the two conditions of the Rayleigh criterion was suggested by other works such as ~\cite{Bhagwat:2016ntk,Forteza:2020hbw}. More recently, \cite{2021arXiv210705609I} also reexamined the Rayleigh criterion and put forward a less restrictive condition based on the distinguishability of the 2-dimensional posteriors for $f_{\ell mn}$ and $\tau_{\ell mn}$ for two modes. In Appendix~\ref{sec:appendix-other-criteria} we present the spectroscopy horizons that result from these various criteria.

\section{Bayesian inference and model comparison}
\label{sec:Bayes}

In this Section we use a fully Bayesian statistical approach to obtain the BH spectroscopy horizons, as an alternative to the Rayleigh criterion used in Section \ref{sec:Rayleigh}. As we briefly explain in the next subsection, our method uses a threshold on the Bayes factor to confirm the presence of one or more secondary modes in the ringdown signal.

\subsection{Bayesian formalism and tests}
\label{sec:bayes_tests}
The posterior probability distribution for a set of parameters $\vartheta$ of a model $\mathcal{M}$, given the detected or assumed data $d$, is
\begin{equation}
	p(\vartheta|d,\mathcal{M}) = \frac{\mathcal{L}(d|\vartheta;\mathcal{M})\pi(\vartheta;\mathcal{M})}{\mathcal{Z}_\mathcal{M}},
	\label{eq:posterior}
\end{equation}
where $\pi(\vartheta;\mathcal{M})$ is the prior probability distribution for the parameters $\vartheta$ which encodes previously known information about the parameters, $\mathcal{L}(d|\vartheta;\mathcal{M})$ is the likelihood of observing the data $d$ given the parameters $\vartheta$ of the model $\mathcal{M}$ and $\mathcal{Z}_\mathcal{M}$ is the evidence, which is the normalization factor of the posterior, given by
\begin{equation}
	\mathcal{Z}_\mathcal{M} = \int \mathcal{L}(d|\vartheta;\mathcal{M})\pi(\vartheta;\mathcal{M}) d\vartheta.
	\label{eq:evidence}
\end{equation}
To quantify the preference for one model $\mathcal{M}_A$ over another model $\mathcal{M}_B$, we use the Bayes factor defined as
\begin{equation}
	\mathcal{B}^{A}_{B} = \frac{\mathcal{Z}_{\mathcal{M}_A}}{\mathcal{Z}_{\mathcal{M}_B}}.
	\label{eq:bayes-factor}
\end{equation}

Assuming GW detectors with stationary Gaussian noise, the likelihood is given by
\begin{equation}
	\mathcal{L}(d|\vartheta;\mathcal{M}) \propto \exp[-\frac{1}{2} \langle d - h(\vartheta; \mathcal{M})| d - h(\vartheta; \mathcal{M})\rangle],
	\label{eq:likelihood}
\end{equation}
where $h(\vartheta, \mathcal{M})$ is the waveform of the model $\mathcal{M}$ with parameters $\vartheta$.

If we assume that the data $d$ contain the detector noise $n$ and two modes, the dominant quadrupolar mode $(2,2,0)$ and a subdominant mode $(\ell, m, n)$, we have $d = d_2$ given by
\begin{equation}
	d_2 = n + h_{220} + h_{\ell m n}.
	\label{eq:data}
\end{equation}

We choose not to use numerical simulation data because we wish to avoid any non-physical numerical noise and the spectral leakage at high frequencies introduced by taking the Fourier transform of the time-domain data with a rectangular window. The spectral leakage can be suppressed by windowing the signal with a smooth ``turn on'', but the QNMs decay very quickly (especially the overtone) and a smooth window could remove the most relevant part of the signal, which is not desirable. For real data, a time domain analysis may be preferable over the frequency domain analysis (see~\cite{Carullo:2019flw,Isi:2020tac,Capano:2021etf,2021arXiv210705609I}), but it is more computationally expensive and the analytical QNMs are a good approximation of the numerical simulation data for our work.

Following Sec.~\ref{sec:Fisher}, we compute the analytical Fourier transform of each QNM using the Flanagan and Hughes prescription~\cite{Flanagan:1997sx}. We use the noise spectral density $S_n(f)$
to generate independent realizations of the noise $n(f)$. In this Section we focus on the case of the LIGO and CE detectors only, due to computational constraints. Out of the remaining future GW detectors considered in this work, we can see from Figure \ref{fig:horizon_rayleigh} that ET and CE will have very similar horizon curves (ET's extends slightly to higher masses, or lower frequencies, see Figures \ref{fig:detector_noise} and \ref{fig:horizon_rayleigh}) and that LISA's horizon will extend far enough for multiple modes to be detected in the ringdown of a supermassive BH binary.

\paragraph{Ringdown models}

The models considered in our analysis, which we suppose to be valid starting at $t = t_{\rm peak} + 10(M_1 + M_2)$, are
\begin{itemize}
	\item $\mathcal{M}_1$, a \emph{single}-mode model with 4 parameters $\vartheta_{1} = \{ A, \phi_{220}, f_{220}, \tau_{220}\}$,
	\item $\mathcal{M}_2$, a \emph{two}-mode model with 8 parameters $\vartheta_{2} = \{A$, $\phi_{220}$, $f_{220}$, $\tau_{220}$, $R_{\ell mn}$, $\phi_{\ell m n}$, $f_{\ell m n}$, $\tau_{\ell m n} \}$,
\end{itemize}
where $R_{\ell m n} = A_{\ell m n}/A$ is the amplitude ratio between a specified subdominant mode $(\ell,m,n)$ and the dominant mode $(2,2,0)$. The global amplitude parameter $A$ is proportional to $M/D_L$ (see eq.~\eqref{eq:qnm}) and all model parameters are defined in the \emph{detector frame}. An agnostic multimode analysis with two and three modes is presented in Sec.~\ref{sec:bayes-multimodes}.

\paragraph{Prior distributions}

As the spectroscopic analysis will look for small contributions to an already detected signal, we consider prior distributions informed by the properties of the binary estimated by the complete inspiral-merger-ringdown (IMR) analysis. The ringdown parameters in the detector frame depend strongly on the final mass $M$ and redshift $z$, which we assume to be estimated with $\pm50\%$ errors at 90\% credibility, that is, within $[M_{\rm min},M_{\rm max}] = [0.5M,1.5M]$ and $[z_{\rm min},z_{\rm max}] = [0.5z,1.5z]$ for each event.\footnote{Note that this interval is not too narrow, as most of the confirmed LIGO events satisfy these conditions for the final mass and redshift~\cite{LIGOScientific:2018mvr,LIGOScientific:2020niy}. The exceptions are GW190909\underline{\space}114149 (mass and redshift), and 11 other events of the O3a catalogue (redshift only). All the events used in the LIGO ringdown analysis have parameters estimated by the IMR analysis inside the interval considered here~\cite{Abbott:2020jks}.}
There is also a weaker dependence on the final spin $a$, or, equivalently, on the binary mass ratio $q$ if we assume circular nonspinning binaries (the only parameter that depends strongly on $q$ is $R_{\ell mn}$). For the cases we consider ($q = 1.5$ and $q = 10$) we fix the corresponding value of $a$ (see Figure \ref{fig:QNMs_a_q} and Table \ref{table:qnm_pars}). Consequently, we choose conservative prior distributions for the model parameters using their expected GR values in the ranges described below:
\begin{itemize}
\item $\pi(A;\mathcal{M}_{1,2})$ is log-uniform in $[A_{\rm min},A_{\rm max}]$ with $A_{\rm min} = A(M_{\rm min},z_{\rm max},a)/10$ and $A_{\rm max} = 10\times A(M_{\rm max},z_{\rm min},a)$,
\item $\pi(\phi_{220};\mathcal{M}_{1,2})$ is uniform in $[0,2\pi]$,
\item $\pi(f_{220};\mathcal{M}_{1,2})$ is log-uniform in $[f^{\rm min}_{220},f^{\rm max}_{220}]$ with $f^{\rm min}_{220} = f_{220}(M_{\rm max},z_{\rm max},a)$ and $f^{\rm max}_{220} = f_{220}(M_{\rm min},z_{\rm min},a)$,
\item $\pi(\tau_{220};\mathcal{M}_{1,2})$ is uniform in $[\tau^{\rm min}_{220},\tau^{\rm max}_{220}]$ with $\tau^{\rm min}_{220} = \tau_{220}(M_{\rm min},z_{\rm min},a)$ and $\tau^{\rm max}_{220} = \tau_{220}(M_{\rm max},z_{\rm max},a)$,

\end{itemize}
for the dominant mode (2,2,0) (used in both models $\mathcal{M}_1$ and $\mathcal{M}_2$), and
\begin{itemize}
\item $\pi(R_{\ell mn};\mathcal{M}_{2})$ is uniform in $[0,0.9]$,
\item $\pi(\phi_{\ell m n};\mathcal{M}_{2})$ is uniform in $[0,2\pi]$,
\item $\pi(f_{\ell mn};\mathcal{M}_{2})$ and $\pi(\tau_{\ell mn};\mathcal{M}_{2})$ are defined in the same way as $\pi(f_{220};\mathcal{M}_{1,2})$ and $\pi(\tau_{220};\mathcal{M}_{1,2})$, but for a \emph{specified} secondary mode $(\ell,m,n)$ in model $\mathcal{M}_2$,
\end{itemize}
where log-uniform priors (i.e., scale-invariant priors) are chosen for the global amplitude and the frequencies.

In order to determine how well models $\mathcal{M}_1$ and $\mathcal{M}_2$ describe the data $d_2 = n + h_{220} + h_{\ell m n}$, we can calculate the deviation between the injected parameters $\vartheta_{\rm inj}^\alpha$ and the posteriors of the estimated parameters $\vartheta^\alpha$ obtained with each model, analogously to a simulation-based calibration procedure \cite{2018arXiv180406788T,2020MNRAS.499.3295R,2021arXiv210705609I}.
This procedure helps identifying inaccurate computations and/or inconsistencies in model implementation.
We present this deviation in units of ``$\sigma$'' by calculating the quantile function $Q(x)$ (associated with the normal distribution) of the integral
\begin{equation*}
x = \int_{-\infty}^{\vartheta_{\rm inj}^\alpha}p(\vartheta^{\alpha}|d_2,\mathcal{M}_{1,2})\ d\vartheta^{\alpha}.
\end{equation*}
For example: if $x \approx 0.159$, then we have $Q(x;p) \approx -1$ (or -1 $\sigma$ deviation), if $x \approx 0.977$, then we have $Q(x;p) \approx 2$ (or +2 $\sigma$ deviation), etc.

\begin{figure}[htb!]
    \centering
    \includegraphics[width = 1.0\linewidth]{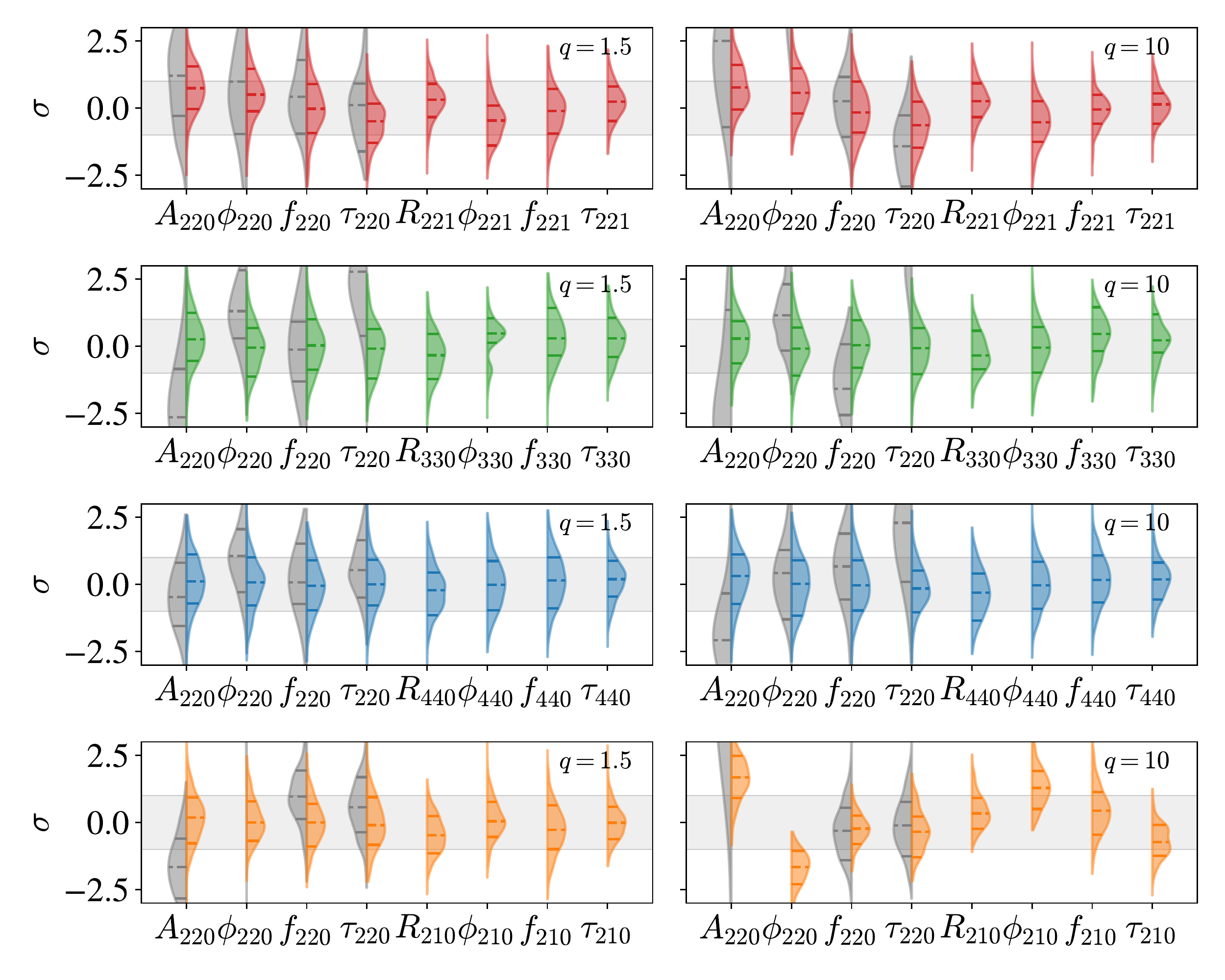}
    \caption{\emph{Parameter estimation bias: Fitting data containing two modes with single-mode model \emph{vs.} two-mode model.} The violin plots show the distribution of the deviation between the injected parameter $\vartheta_{\rm inj}^\alpha$ and the posterior probability of $\vartheta^\alpha$ from the data $ d_2 = n + h_{220} + h_{\ell m n}$, see main text for more details. The results of each panel were obtained with 500 signals $d_2$ with different parameters and independent noise realizations (see main text for details). The left-sided violins (gray) show the deviation of each parameter of the dominant mode estimated with the single-mode model $\mathcal{M}_1$ and the colored right-sided violins show the deviation of all the parameters of the correct two-mode model $\mathcal{M}_2$, considering the subdominant modes $(2,2,1)$ (red), $(3,3,0)$ (green), $(4,4,0)$ (blue) and $(2,1,0)$ (orange). Model $\mathcal{M}_1$ neglects the subdominant mode and produces results that are more biased and have a larger spread of deviations.}
    \label{fig:2modes-parameters}
\end{figure}

In Figure~\ref{fig:2modes-parameters} we show these results in the violin plots. For each case we generated 500 signals $d_2$, with independent noise realizations and $50$ final masses $\times$ 10 redshifts distributed around the spectroscopy horizon for each subdominant mode (see Figure~\ref{fig:horizon_bayes_2modes}).
The left-sided violins (in gray) show the distributions of deviations for each injected parameter of the dominant mode $(2,2,0)$ with the posteriors estimated with the single-mode model $\mathcal{M}_1$. The colored right-sided violins show the same distributions for all of the parameters using the correct two-mode model $\mathcal{M}_2$. Model $\mathcal{M}_1$ estimates the parameters of the fundamental mode with larger bias and a greater spread of deviations to compensate for neglecting the subdominant mode. The bias is not substantial because the subdominant mode can be neglected in half of the signals, which are outside the spectroscopy horizon.\footnote{The exception are the grey violin plots for $\phi_{220}$ associated with (2,2,0)+(2,2,1) data for $q = 10$ and with (2,2,0)+(2,1,0) data for both $q = 1.5$ and $q = 10$, which are outside the axis limits and represent the largest bias ($> 6 \sigma$).} Therefore, both models produce posteriors compatible with the injected parameters in this distribution, but model $\mathcal{M}_1$ will have significant bias when the secondary mode is strong.

We use the Bayes factor $\mathcal{B}_1^2$ to quantify the preference for model $\mathcal{M}_2$ over model $\mathcal{M}_1$ in a given signal. Additionally, we choose a detectability threshold of $\ln \mathcal{B}_1^2 > 8$, that is, the evidence $\mathcal{Z}_2$ of the two-mode model must be approximately $3000$ greater than the evidence $\mathcal{Z}_1$ of the single-mode model for a confident claim of detection of a secondary mode.

\begin{figure}[htb!]
	\centering
	\includegraphics[width = 1.0\linewidth]{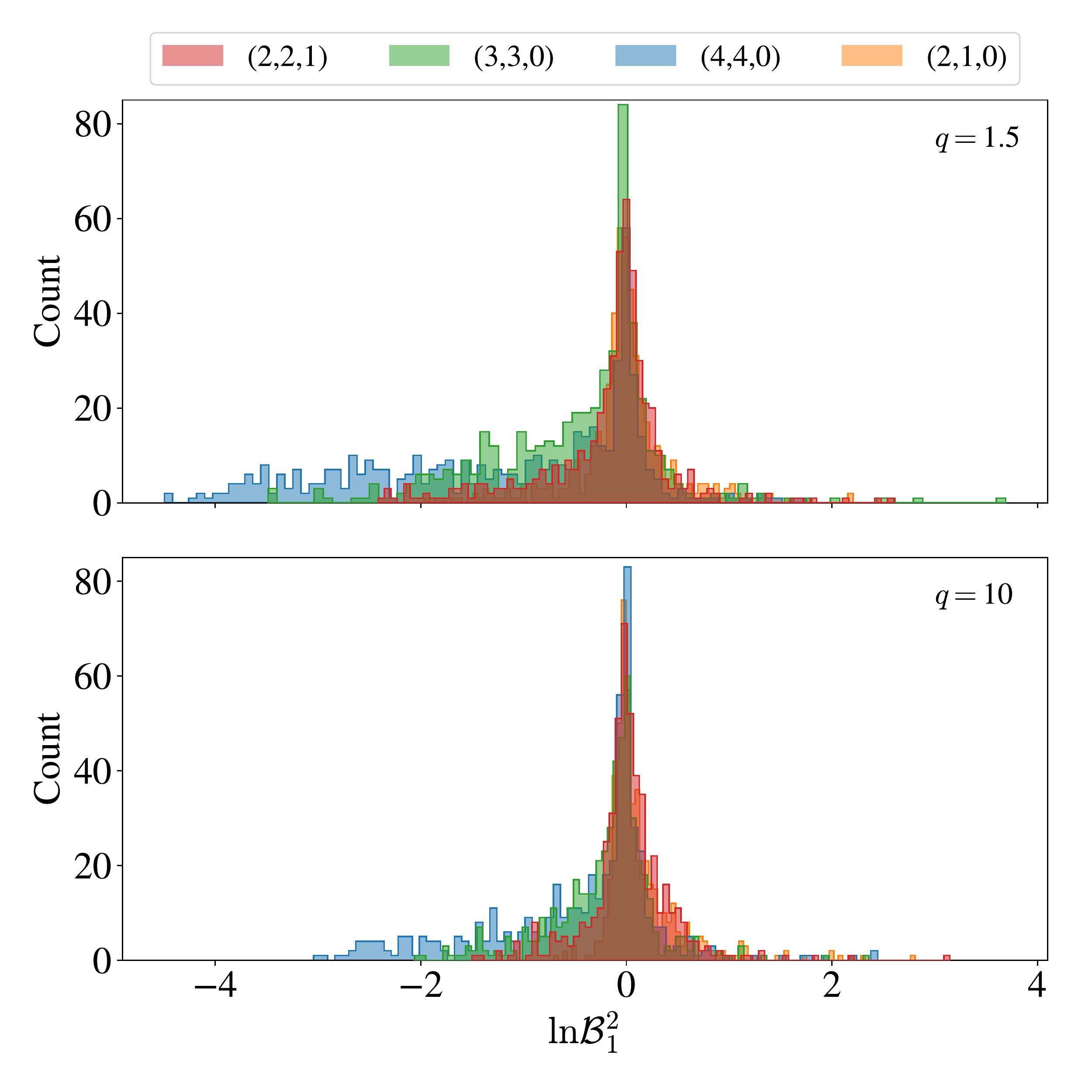}
	\caption{\emph{Bayes factor threshold for confident detections: Looking for two modes in a single-mode signal.} The histograms show the natural logarithm of the Bayes factor $\mathcal{B}_{1}^{2}$ in favor of the two-mode model $\mathcal{M}_2$ over the single-mode model $\mathcal{M}_1$, for 2000 signals containing single-mode data, $d_{1} = n + h_{220}$ with LIGO noise realizations. The colors indicate the prior of the subdominant mode in each case: $(2,2,1)$ (red), $(3,3,0)$ (green), $(4,4,0)$ (blue) and $(2,1,0)$ (orange), see main text for more details. For each prior we generated 500 signals with $M \in [1, 5\times10^3]M_\odot$ and $z \in [10^{-2}, 1]$. For most signals the single-mode model is favored ($\mathcal{B}_1^2 \lesssim 1$), as expected, but we find 5 outliers with $10 \lesssim \mathcal{B}_1^2 \lesssim 55$. We require $\ln \mathcal{B}_1^2 > 8$ ($\mathcal{B}_1^2 \gtrsim 3000$) for claiming a confident detection of a secondary mode in a ringdown signal.}
	\label{fig:histogram}
\end{figure}

This choice is supported by the results presented in Figure~\ref{fig:histogram}, where we compute the Bayes factor $\mathcal{B}_1^2$ using data that contain only the dominant mode $d_{1} = n + h_{220}$ (to verify that $\ln \mathcal{B}_1^2$ is never greater than $8$ in this case). As expected, most events have $\ln \mathcal{B}_1^2 \lesssim 0$ (or $\mathcal{B}_1^2 \lesssim 1$). We do not find any cases with  $\ln \mathcal{B}_1^2 > 4$ (or $\mathcal{B}_1^2 \gtrsim 55$), but we caution that 5 out of the 2000 simulated signals ($\sim 0.3\%$ of all cases) had $10 \lesssim \mathcal{B}_1^2 \lesssim 55$, which is considered to be ``strong'' evidence according to the widely used scale by Kass and Raftery \cite{doi:10.1080/01621459.1995.10476572}. This indicates that such generic scales may not be suitable for different types of data (and noise), and should preferentially be avoided in the analysis of GW data if a more detailed analysis is possible. 

In this and the following Sections, the evidences were calculated using \textsc{PyMultiNest}~\cite{Feroz:2008xx,Feroz:2013hea,Buchner:2014nha} with 500 live points in each computation. A spot check with 1000 live points showed that the evidence changed by approximately 6\%.

\subsection{Bayesian model comparison: looking for a secondary mode}
\label{sec:bayes-two-mode}
Recent multi-mode analyses of the ringdown of some GW detections found that the inclusion of subdominant modes decreases the errors in the inferred parameters of the final BH~\cite{Isi:2019aib,Abbott:2020tfl,Abbott:2020jks,2021arXiv210705609I}. However, the Bayes factors in favor of models with additional modes are close to 1.

In~\cite{Bustillo:2020tjf} the authors proposed the use of the ringdown part of the waveform approximants to probe the Kerr nature and found a high statistical evidence of $\ln \mathcal{B}\sim 6.5$ in favor of their model with respect to their best QNM model for GW150914. However, their model does not provide evidence for the individual detection of multiple QNMs.
More recently, a new analysis of the ringdown of the most massive event GW190521~\cite{Capano:2021etf} had $\mathcal{B}_1^2\sim 40$ ($\ln\mathcal{B}_1^2\sim 3.7$). This is the highest evidence for a secondary mode presented so far.

We emphasize that the Bayes factor depends on the choice of priors and results from different works should not be directly compared without this context.  In our analysis we use Figure~\ref{fig:histogram} to determine whether a given value for the Bayes factor provides ``strong'' evidence for each set of priors. We choose here the detectability threshold $\ln\mathcal{B}_1^2 = 8$; lower thresholds would result in larger spectroscopy horizons, albeit with lower significance.

\begin{figure}[htb!]
	\centering
	\includegraphics[width = 1.0\linewidth]{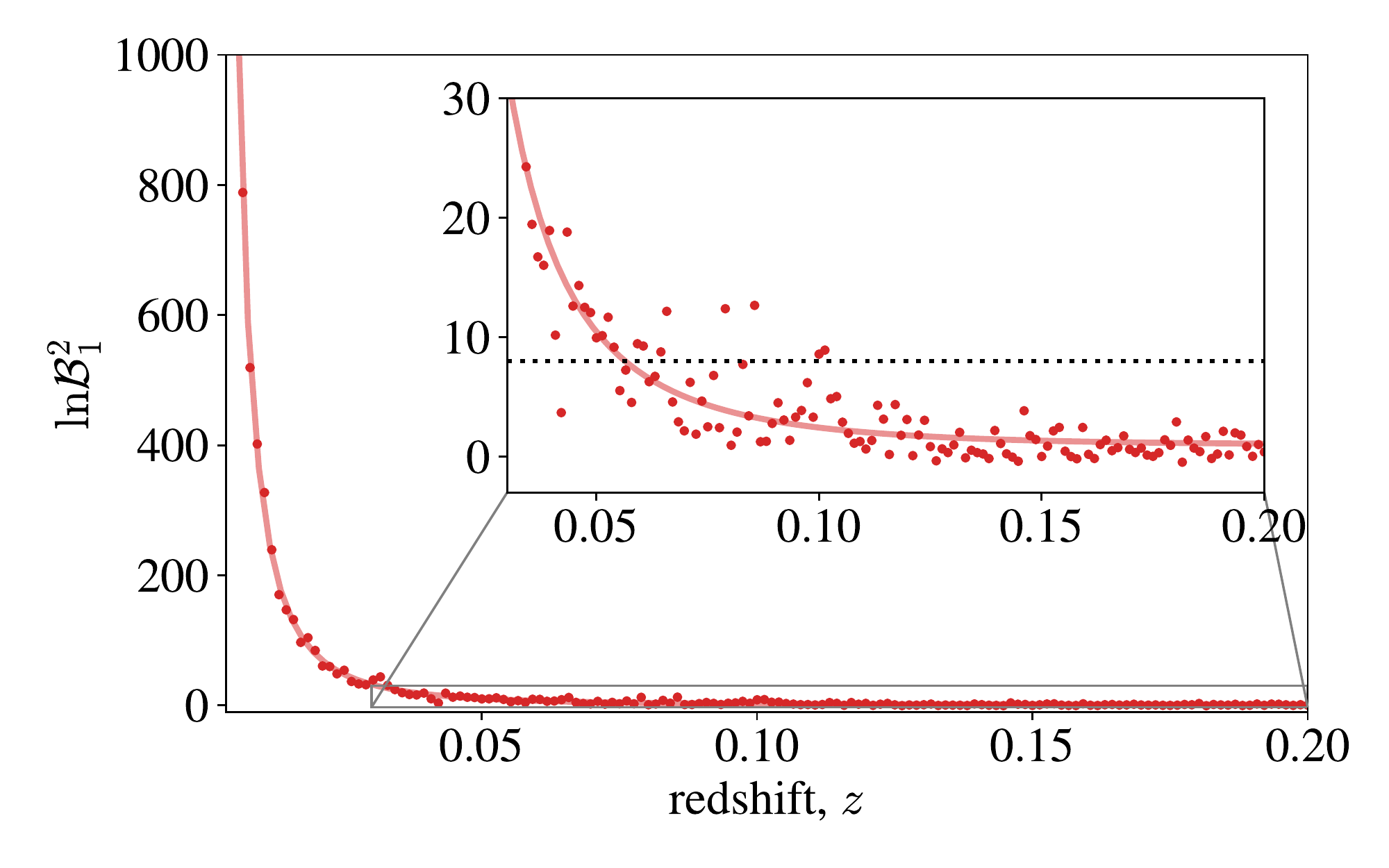}
	\caption{Bayes factor of the two-mode model $\mathcal{M}_2$, here with the overtone $(2,2,1)$ prior for the subdominant mode, over the single-mode model $\mathcal{M}_1$ as a function of redshift. There are 150 simulated signals with data $d_2 = h_{220} + h_{221} + n$, with fixed $M = 156M_\odot$, mass ratio $q = 1.5$ and independent LIGO noise realizations. The errors in the Bayes factors, as estimated by MultiNest, are smaller than the points. The scattering of the points is due to the independent noise realizations, but it is averaged out in the computation of the BH spectroscopy horizon, see Figure \ref{fig:horizon_bayes_2modes}. The continuous curve is a Laurent polynomial $c_1 + c_2z^{-1} + c_3z^{-2}$ fitted to the points. The redshift at the detectability threshold $\ln\mathcal{B}_1^2 = 8$ is determined using the fitted curve.}
	\label{fig:bayes_redshift_fit}
\end{figure}

The calculation of Bayes factors is very computationally expensive, and the time to find the root $\ln\mathcal{B}_{1}^{2} - 8 = 0$ with independent noise realizations can be very long if one uses standard root-finding algorithms.
However, for a fixed mass, the Bayes factor is approximately monotonic as a function of redshift. In Figure~\ref{fig:bayes_redshift_fit} we show an example of $\ln\mathcal{B}_1^2$ as a function of the redshift for many simulated signals with data $d_2 = h_{220} + h_{221} + n$ for events with fixed final mass and mass ratio at different distances. We find that $\ln\mathcal{B}_1^2$ is well approximated by a Laurent polynomial fit, which allows us to find the redshift at which $\ln\mathcal{B}_{1}^{2} - 8 = 0$ for the fixed mass. This will be the redshift $z^{\rm spec, B}_{\ell mn}$ of the BH spectroscopy horizon for that mass.

To find the BH spectroscopy horizon using the Bayes factor threshold for a given GW detector, we choose several masses spaced log-uniformly in the range of interest. For each mass, initial samplings of  $\ln\mathcal{B}_{1}^{2}$ at different distances provide a rough estimate of $z^{\rm spec, B}_{\ell mn}$ for which $\ln\mathcal{B}_{1}^{2} = 8$. This result is then refined using a Laurent polynomial fit with 10 (30) points chosen within $\pm 50\%$ of the rough estimate for LIGO (CE). Different thresholds could also be used with our method.

\begin{figure}[htb!]
	\centering
	\includegraphics[width = 1.0\linewidth]{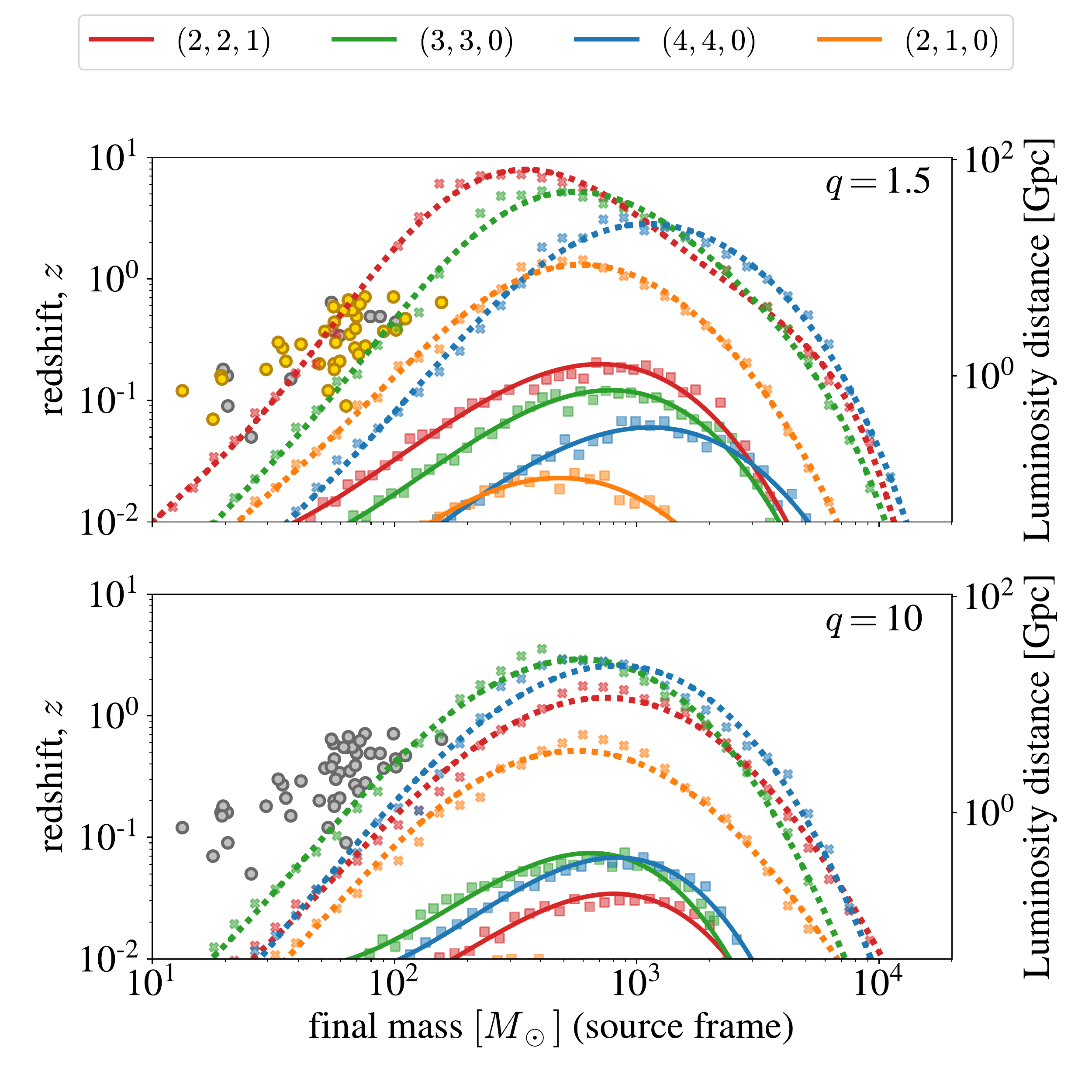}
	\caption{CE (crosses and dotted curves) and LIGO (squares and solid curves) BH spectroscopy horizons obtained with signals containing two modes, $d_2 = n + h_{220} + h_{\ell mn}$, by requiring Bayes factors $\ln \mathcal{B}_{1}^{2} >8$ for the subdominant modes $(2,2,1)$ (red), $(3,3,0)$ (green), $(4,4,0)$ (blue) and $(2,1,0)$ (orange). The curves are polynomials fitted to the points, with coefficients given in Tables ~\ref{table:fits_2_modes} and ~\ref{table:fits_2_modes_CE}. Circles show detections from GWTC-1 and GWTC-2; yellow circles are compatible with the mass ratio in each case, whereas gray circles are not.}
	\label{fig:horizon_bayes_2modes}
\end{figure}

In Figure~\ref{fig:horizon_bayes_2modes} we show the LIGO and CE BH spectroscopy horizons up to which $\ln \mathcal{B}_{1}^{2} >8$ for signals with data $d_2$ containing two modes. We fit the horizons with 3rd to 7th degree polynomials; all coefficients are given in Tables ~\ref{table:fits_2_modes} (LIGO) and \ref{table:fits_2_modes_CE} (CE). In the $q=1.5$ case, for lower masses, the first overtone $(2,2,1)$ has the largest horizon, and the second largest is the $(3,3,0)$ mode followed by the $(2,1,0)$. For higher masses, the horizon of the $(4,4,0)$ is slightly larger. For $q=10$ the $(3,3,0)$ and $(4,4,0)$ modes have larger horizons than the $(2,2,1)$ mode and the $(2,1,0)$ mode has consistently the smallest horizon.
The qualitative features are similar to the BH spectroscopy horizons obtained with the Rayleigh criterion (see Figure~\ref{fig:horizon_rayleigh}).

The BH spectroscopy horizon distance depends on the amplitude, the frequency and the damping time of the modes. As we can see in Table~\ref{table:qnm_pars} for $q=1.5$, the overtone has a much higher amplitude (at least a factor 5) than the higher harmonics and therefore it has the largest horizon. For $q=10$, the amplitudes are much closer, and the overtone horizon is smaller than the $(3,3,0)$ and $(4,4,0)$ horizons, even though the amplitude of the overtone is still slightly larger than the amplitude of the higher harmonics. This shows that the difference between the frequency of the dominant and subdominant modes is more relevant than the difference between decay times, as the $(3,3,0)$ and $(4,4,0)$ modes have damping times very close to the damping time of the $(2,2,0)$ mode and very different frequencies, and there is an opposite relation between the $(2,2,1)$ and $(2,2,0)$ modes. Furthermore, for $q=10$ the amplitude of the $(4,4,0)$ mode is less than half of the amplitude of the $(3,3,0)$ mode, but their horizons peak at approximately the same distance. This happens because the higher frequency of the $(4,4,0)$ mode remains in the detector band for more massive BHs.

Moreover, the $(2,1,0)$ mode always has a higher amplitude than the $(4,4,0)$ mode and for $q=10$ it has an amplitude comparable with the $(2,2,1)$ and $(3,3,0)$ modes, but its horizon is always much smaller. This is because the frequency \emph{and} damping time of the $(2,1,0)$ mode are very close to the dominant mode's values. Therefore, the difference in damping times is also very important, as the overtone horizon is much larger than the $(2,1,0)$ horizon.

These results confirm the importance of \emph{both} $\Delta f_{220,\ell mn}$ and $\Delta \tau_{220,\ell mn}$ used in the Rayleigh criterion~\eqref{eq:rayleigh_both}, however the horizons obtained using the Bayes factor threshold are much larger from the ones obtained using the Fisher matrix analysis and the Rayleigh criterion. For each secondary mode, the maximum horizon distance calculated using the Bayes factor threshold is at least a few times larger than the maximum horizon distance obtained with the Rayleigh criterion. The comparison between the methods is discussed in Sec.~\ref{sec:discussion}.

On the horizon $z^{\rm spec, B}_{\ell mn}$, the SNR of the subdominant modes is larger than approximately 5 in all cases. The posterior probability distributions for all parameters are informed by the data and exclude zero amplitude of the secondary mode at the 90\% credibility level (we examine one example in detail in Figure \ref{fig:fisher_parameters_gw190521} below).

We can use our results together with the current rates for binary BH mergers to estimate the rate of events expected within the BH spectroscopy horizons. Recently, \cite{2021ApJ...913L...7A} quoted the rate of binary mergers with primary mass $45 M_{\odot} < M_1 < 100 M_{\odot}$ as $0.70^{+0.65}_{-0.35}\ {\rm Gpc}^{-3}\ {\rm yr}^{-1}$. Using the comoving volume at $z^{\rm spec, B}_{221}$ for $q = 1.5$ and final mass $M = 156.3 M_{\odot}$ (similar to GW190521), we find event rates $0.03 - 0.10\ {\rm yr}^{-1}$ for LIGO and $(0.6 - 2.4) \times 10^3\ {\rm yr}^{-1}$ for CE.

The horizons are smaller for lower masses and for other subdominant modes; optimal inclinations and sky locations could increase the horizons by less than a factor 2. We will  have to be lucky to see such an event with LIGO, but there should be no lack of detections when the 3G detectors are operational.

\begin{table}[!htbp]
\caption{Coefficients used in the LIGO BH spectroscopy horizon fits shown in Figure ~\ref{fig:horizon_bayes_2modes}, obtained with the Bayes factor condition $\ln \mathcal{B}_1^2 = 8$. The fits have the form $\ln z^{\rm spec, B}_{\ell mn} = a_0 + a_1 \ln M + a_2 (\ln M)^2 + a_3 (\ln M)^3$, where $z^{\rm spec, B}_{\ell mn}$ is the redshift at the horizon distance and $M$ the black hole final mass. These fits are only valid for the mass intervals shown in Figure~\ref{fig:horizon_bayes_2modes}.}
\begin{ruledtabular}
\begin{tabular}{c c c c c}
coefficient & $(2,2,1)$ & $(3,3,0)$ & $(4,4,0)$ & $(2,1,0)$  \\ \hline\hline
\multicolumn{5}{c}{$q = 1.5$}\\ \hline
$a_0$ & -0.6685 & -0.7298 & -0.5887 & -0.0022 \\
$a_1$ & 4.0504 & 4.6084 & 3.8802 & -1.3164	\\
$a_2$ & -6.8045 & -8.3415 & -7.1914 & 7.0597 	\\
$a_3$ & 1.2680 &  2.3177 &  1.3209 & -11.0706	\\ \hline\hline

\multicolumn{5}{c}{$q = 10$}\\ \hline
$a_0$ & -0.8621 & -0.9526 & -1.1328 & \\
$a_1$ & 5.7614 & 6.1710 & 7.8589 & \\
$a_2$ &-11.6508 & -12.1380 & -16.9297 &\\
$a_3$ & 4.8845 & 5.3732 & 9.4642 &\\
\end{tabular}
\end{ruledtabular}
\label{table:fits_2_modes}
\end{table}

\begin{table}[!htbp]
\caption{Same as Table \ref{table:fits_2_modes}, but for the CE detector.}
\begin{ruledtabular}
\begin{tabular}{c c c c c}
coefficient & $(2,2,1)$ & $(3,3,0)$ & $(4,4,0)$ & $(2,1,0)$  \\ \hline\hline
\multicolumn{5}{c}{$q = 1.5$}\\ \hline
$a_0$ & -0.2331 & -0.2081  & -0.1342 & -0.1491 \\
$a_1$ & 3.3135 & 3.1694 & 2.1851 &  2.3180 \\
$a_2$ & -18.5618 & -19.3023 & -14.4192 & -14.5241 \\
$a_3$ & 51.7728 & 59.5351 & 48.7304 & 46.3409 \\
$a_4$ & -75.6249 & -98.0044 & -88.7223 & -79.3446 \\
$a_5$ & 57.1219 & 84.0752 & 84.6849 & 71.1739 \\
$a_6$ & -19.7984 & -32.0157 & -36.0500 & -28.7501  \\ \hline\hline

\multicolumn{5}{c}{$q = 10$}\\ \hline
$a_0$ & 0.1135 & 0.0647 & -0.0107 & 0.1703 \\
$a_1$ & -1.5090 & -1.0693 & -0.3054 & -1.9806 \\
$a_2$ & 5.8306 & 4.1535 & 1.4965 & 7.0495 \\
$a_3$ & -6.9124 & -3.9086 & 0.1276 & -8.0004 \\
$a_4$ & -0.07017 & -1.6608 & -4.3371 & -0.1510
\end{tabular}
\end{ruledtabular}
\label{table:fits_2_modes_CE}
\end{table}

\subsection{Multimode LIGO BH spectroscopy horizons}
\label{sec:bayes-multimodes}

The ringdown of a Kerr black hole is actually a superposition of infinite modes. The two-mode approximation is valid when the tertiary mode (and all of the higher modes) are too weak to be detected in the signal.\footnote{Additionally, the $(\ell,m) \ne (2,2)$ harmonics cannot be detected in events viewed face-on ($\iota = 0$) and harmonics with odd $m$ are not excited in non-spinning circular binaries with mass ratio $q=1$.}

In this section we consider the constraints imposed by a more realistic signal, where we assume that the data contains the noise, the dominant mode and the four most relevant subdominant modes~\cite{Cotesta:2018fcv}, with their respective amplitudes informed by numerical relativity simulations. That is, we have $d = d_5$ given by
\begin{equation}
    d_5 = n + h_{220} + h_{221} + h_{330} + h_{440} + h_{210}.
    \label{eq:data-5modes}
\end{equation}
The models considered in the multimode analysis are
\begin{itemize}
    \item $\mathcal{M}_{1}$, the same \emph{single}-mode model defined in Section \ref{sec:bayes_tests},
    \item $\mathcal{M}_{2}$, a \emph{two}-mode model with one unspecified mode $(\ell_1,m_1,n_1)$ and 8 parameters $\vartheta_{2} = \{A$, $\phi_{220}$, $f_{220}$, $\tau_{220}$, $R_{\ell_1 m_1n_1}$, $\phi_{\ell_1 m_1 n_1}$, $f_{\ell_1 m_1 n_1}$, $\tau_{\ell_1 m_1 n_1} \}$,
    \item $\mathcal{M}_{3}$, a \emph{three}-mode model with two unspecified modes $(\ell_1,m_1,n_1)$ and $(\ell_2,m_2,n_2)$ and 12 parameters $\vartheta_{3} = \{A$, $\phi_{220}$, $f_{220}$, $\tau_{220}$, $R_{\ell_1 m_1 n_1}$, $\phi_{\ell_1 m_1 n_1}$, $f_{\ell_1 m_1 n_1}$, $\tau_{\ell_1 m_1 n_1}$, $R_{\ell_2 m_2 n_2}$, $\phi_{\ell_2 m_2 n_2}$, $f_{\ell_2 m_2 n_2}$, $\tau_{\ell_2 m_2 n_2}\}$.
\end{itemize}
The two-mode model $\mathcal{M}_{2}$ considered in this Section contains the same number of parameters as the model $\mathcal{M}_{2}$ defined in Section \ref{sec:bayes_tests}, but here it has a much broader prior. We are now looking for an unspecified subdominant mode and the priors need to allow the secondary mode to be any of the most relevant subdominant modes $(2,2,1)$, $(3,3,0)$, $(4,4,0)$ and $(2,1,0)$ (following the prescription described in Section ~\ref{sec:bayes-two-mode}). The same restrictions are applied to the priors of the secondary and tertiary modes in the three-mode model $\mathcal{M}_3$.

\begin{figure}[htb!]
    \centering
    \includegraphics[width = 1.0\linewidth]{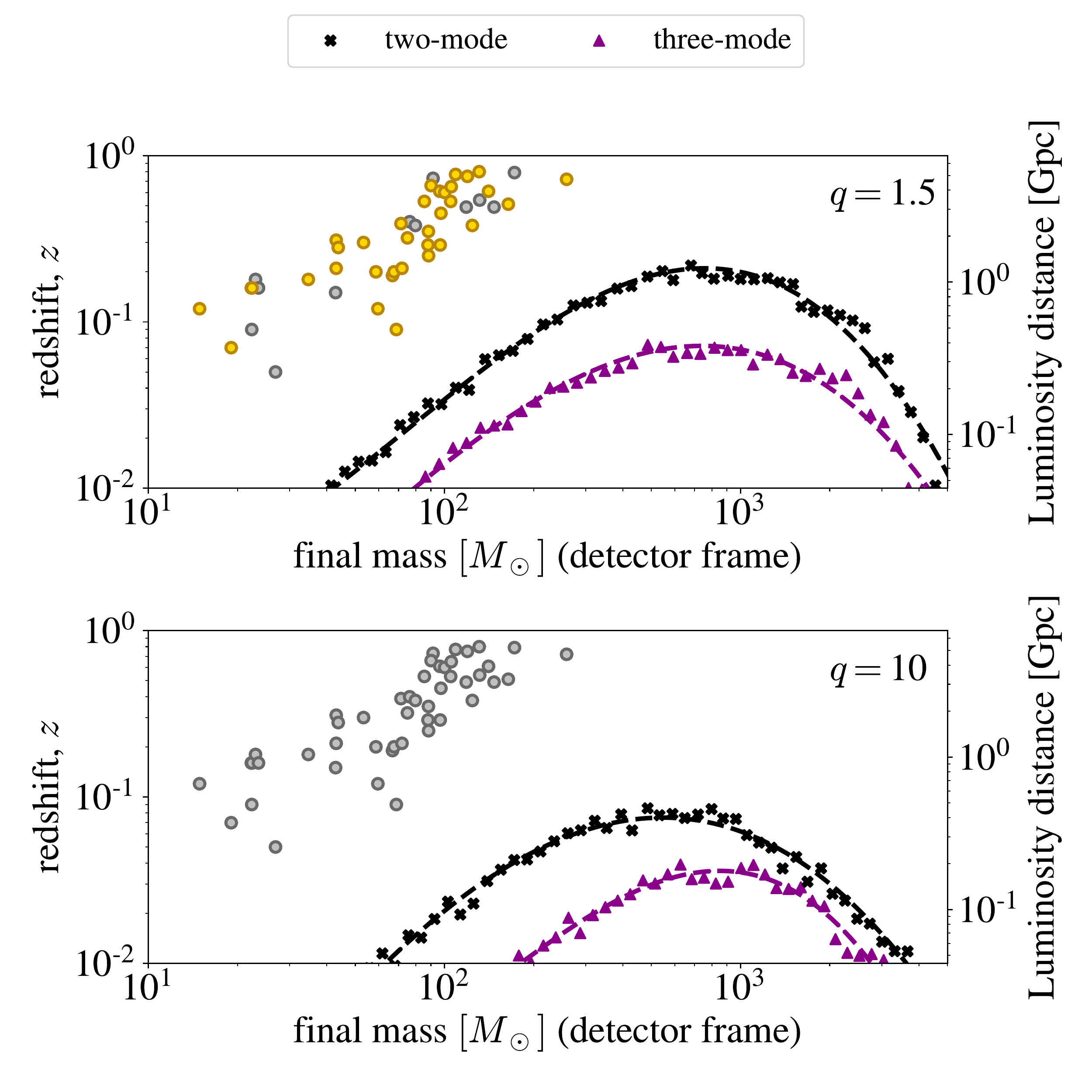}
    \caption{Multimode BH spectroscopy horizons for LIGO, using signals containing five modes, $d_5 = n + h_{220} + h_{221} + h_{330} + h_{440} + h_{210}$, by requiring Bayes factors $\mathcal{B}_{n-1}^{n} >8$ for $n = 2$ (black) and $n=3$ (purple). Here we are looking for the most relevant, but unspecified, secondary and tertiary modes in the five-mode signal. The dashed curves are 3rd degree polynomials fitted to the horizons.}
    \label{fig:horizon_bayes_multi}
\end{figure}

In Figure~\ref{fig:horizon_bayes_multi} we show the multimode BH spectroscopy horizons for LIGO obtained with signals $d_5$ containing five modes, by requiring Bayes factors $\mathcal{B}_{n-1}^{n} >8$, for $n = 3$ and $n = 2$ modes, where the Bayes factors $\mathcal{B}_{n-1}^{n}$ are relative to the models $\mathcal{M}_n$ and $\mathcal{M}_{n-1}$.
The dashed curves are a 3rd degree polynomial fitted to the horizon points, with coefficients given in Table~\ref{table:fits_N_modes}.
As expected, the two-mode horizons are larger than the three-mode horizons. The $q=1.5$ horizons are also larger than the $q=10$ horizons, as expected. This happens because, for the same final mass, nonspinning binaries with more asymmetric initial masses emit less energy in the form of gravitational waves~\cite{Kamaretsos:2011um,London:2014cma,Cotesta:2018fcv}, and the mode amplitudes are smaller than the amplitudes of more symmetric systems (see  Table~\ref{table:qnm_pars}).

\begin{table}[!htbp]
\caption{LIGO Bayes factor $\ln \mathcal{B}_1^2 = 8$ spectroscopy horizons fitting coefficients, $\ln z_n^{\rm spec, B} = a_0 + a_1 \ln M + a_2 (\ln M)^2 + a_3 (\ln M)^3$, where $z_n^{\rm spec, B}$ is the redshift at the horizon and $M$ the black hole final mass. These fits are only valid for the mass intervals of the points in the Figure~\ref{fig:horizon_bayes_multi}.}
\begin{ruledtabular}
\begin{tabular}{c c c}
coefficient & 2-mode model & 3-mode model\\ \hline\hline
\multicolumn{3}{c}{$q = 1.5$}\\ \hline
$a_0$ & -0.4238 & -0.3022 \\
$a_1$ & 2.2212 & 1.3808 \\
$a_2$ & -2.3470 & -0.4939 \\
$a_3$ & -2.2267 & -3.9584\\ \hline\hline

\multicolumn{3}{c}{$q = 10$}\\ \hline
$a_0$ & -0.1121 & -0.7089 \\
$a_1$ & -0.2280 & 4.5011  \\
$a_2$ & 3.7164 & -8.1516 \\
$a_3$ & -7.2915 &  1.6289  \\
\end{tabular}
\end{ruledtabular}
\label{table:fits_N_modes}
\end{table}

\begin{figure}[htb!]
    \centering
    \includegraphics[width = 1.0\linewidth]{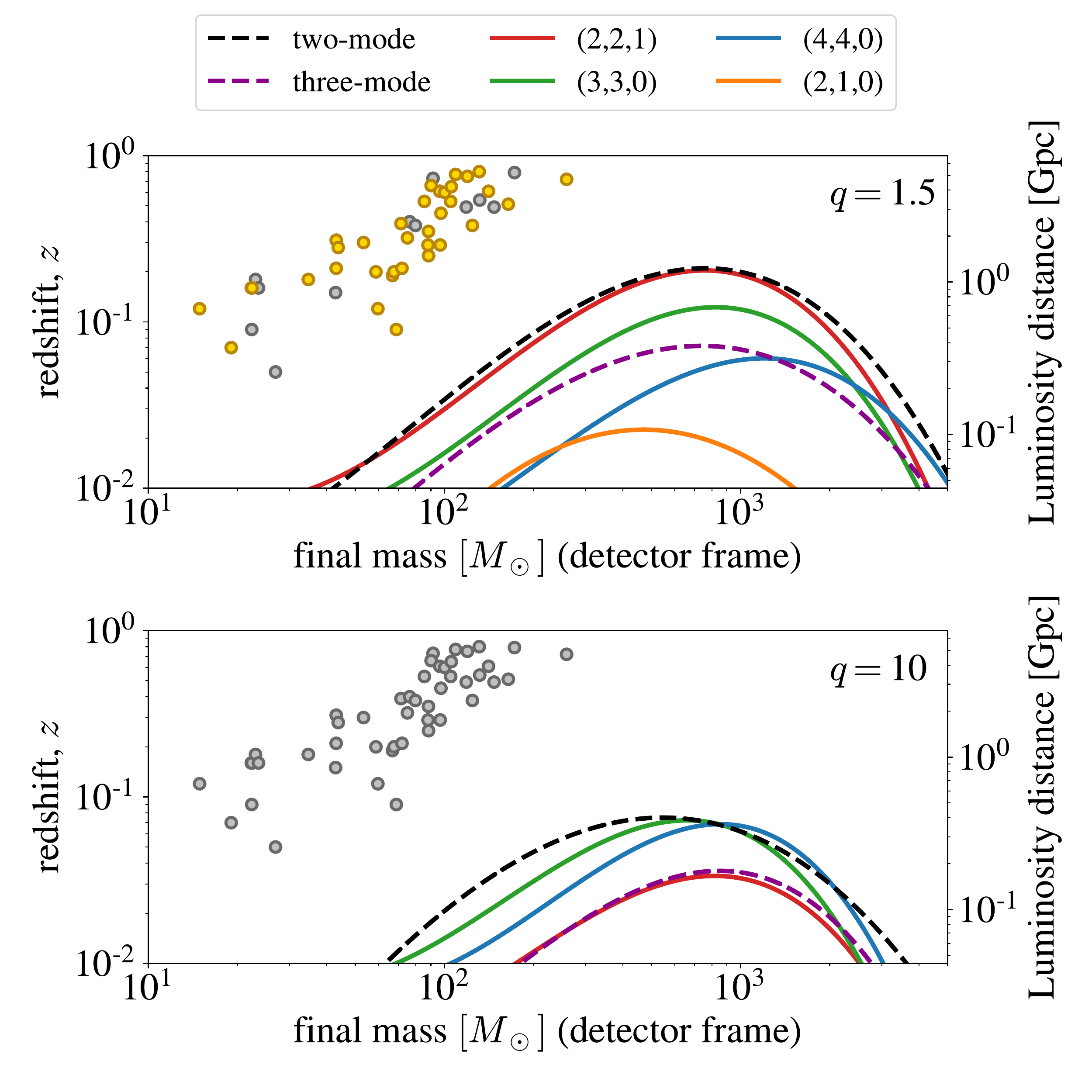}
    \caption{Comparison between the LIGO spectroscopy horizons presented in Figure \ref{fig:horizon_bayes_2modes},  shown with solid lines, and the LIGO multimode spectroscopy horizons presented in Figure \ref{fig:horizon_bayes_multi}, shown with dashed lines. The (unspecified) two-mode horizon follows the largest (specified) horizon for each mass; the three-mode horizon is smaller than the second largest horizon.}
    \label{fig:horizon_bayes_multi_vs_2modes}
\end{figure}

Figure~\ref{fig:horizon_bayes_multi_vs_2modes} compares the LIGO spectroscopy horizons computed in Section~\ref{sec:bayes-two-mode} with data containing only two modes (shown in Figure \ref{fig:horizon_bayes_2modes}), and the LIGO multimode horizons presented in Figure \ref{fig:horizon_bayes_multi}. For $q=1.5$, the two-mode horizon is equivalent to the $(2,2,1)$ horizon. However, the three-mode horizon is slightly smaller than the $(3,3,0)$ horizon, which is the second largest. This may be due to the penalization of a high number of parameters in the Bayes factor. For $q=10$ the two-mode horizon is compatible with the $(3,3,0)$ and the $(4,4,0)$ horizons, all of which peak at approximately the same distance. Again, the three-mode horizon is smaller than the second largest horizon and it even seems to be compatible with the $(2,2,1)$ horizon.

\begin{figure}[htb!]
    \centering
    \includegraphics[width = 1.0\linewidth]{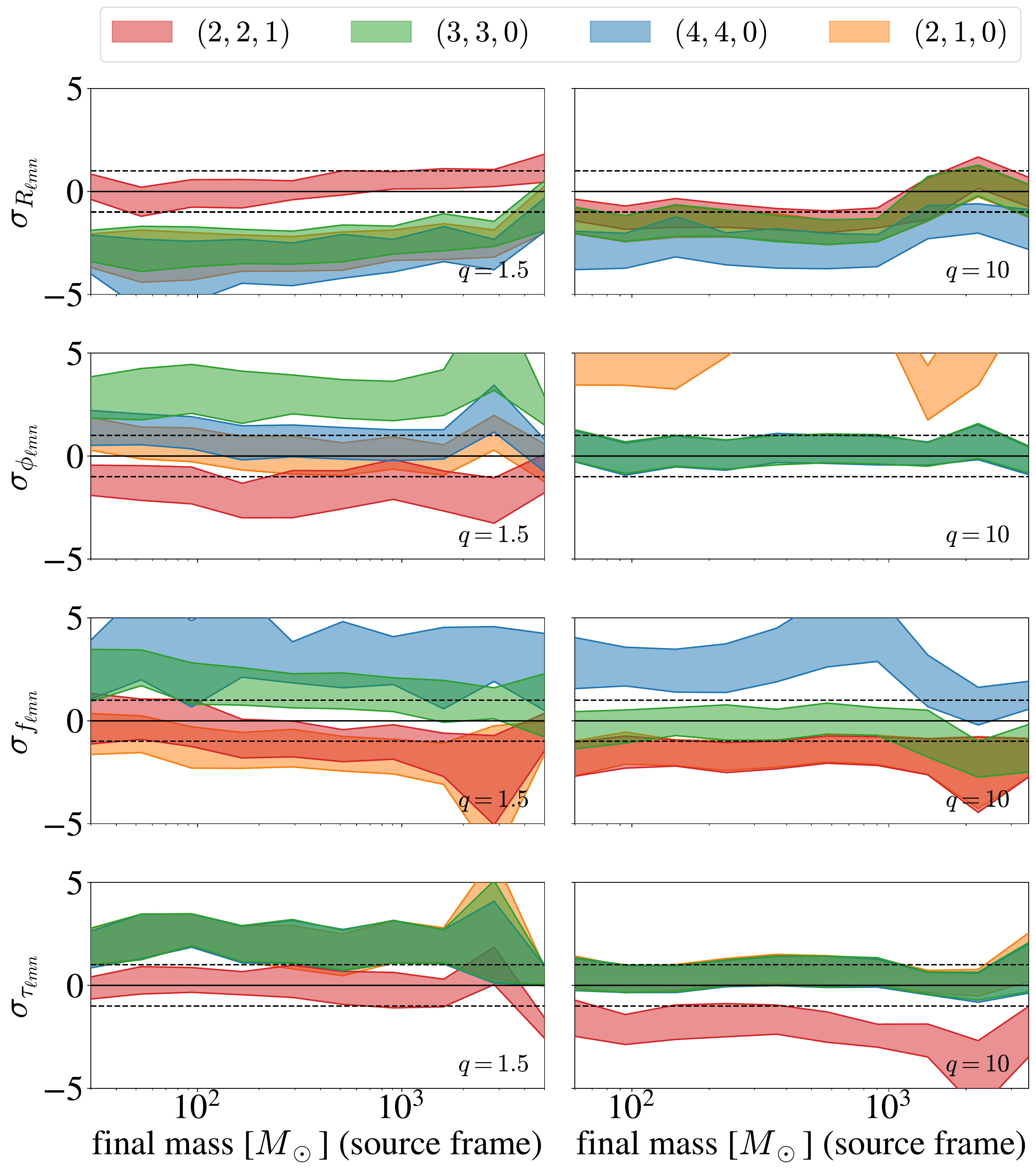}
    \caption{\emph{Identifying the secondary mode:} deviations between the posteriors of parameters $\vartheta^\alpha$ of the unspecified secondary mode in model $\mathcal{M}_2$ and the injected parameters $\vartheta_{\rm inj}^\alpha$ of each of the four subdominant modes in the data $ d_5 = n + h_{220} + h_{221} + h_{330} + h_{440} + h_{210}$. The colored bands show the $\pm 1 \sigma$ highest probability density region of the distributions of the deviations in each case. For $q = 1.5$ (left), the $(2,2,1)$ is the secondary mode detected; for $q = 10$, it is the (3,3,0) mode.}
    \label{fig:errors_pars_2_modes}
\end{figure}

To determine which are the secondary and tertiary modes detected on the two- and three-mode horizons we compute the deviation of each estimated parameter relative to all of the injected modes (following the same procedure detailed in Section~\ref{sec:bayes_tests}). In Figure~\ref{fig:errors_pars_2_modes} we show the deviations calculated \emph{on} the two-mode horizon, that is, for each mass the redshift is fixed to the redshift of the horizon, $z^{\rm spec, B}_{n=2}$. We consider the mass ranges where $z^{\rm spec, B}_{n=2} > 10^{-2}$, as in Figure~\ref{fig:horizon_bayes_multi}. For each horizon we select 10 final masses spaced log-uniformly, and for each mass we generate 100 signals with independent noise realizations.
The colored bands show the $\pm 1 \sigma$  highest probability density region of the deviations.

As the secondary mode is unspecified, and the priors are broad enough to accommodate any of the four most relevant subdominant modes, the secondary mode identification can change over the two-mode horizon. For $q = 1.5$, the secondary mode is compatible with the $(2,2,1)$ mode in almost the entire mass range, as expected from Figure~\ref{fig:horizon_bayes_multi_vs_2modes}. For the highest masses $M\gtrsim 10^3 M_{\odot}$ (lowest frequencies), the secondary mode becomes compatible with the $(3,3,0)$ mode, which has a higher frequency and therefore remains longer in the detector band. This trend is especially clear from the deviations in the amplitude ratio $R_{\ell mn}$ and damping time $\tau_{\ell mn}$. For $q=10$, Figure~\ref{fig:horizon_bayes_multi_vs_2modes} is not enough to identify the secondary mode,  but Figure~\ref{fig:errors_pars_2_modes} allows us to identify it as the $(3,3,0)$ mode for most of the mass range and the $(4,4,0)$ mode for the highest masses, similarly to the transition in the low mass ratio case.

\begin{figure}[htb!]
    \centering
    \includegraphics[width = 1.0\linewidth]{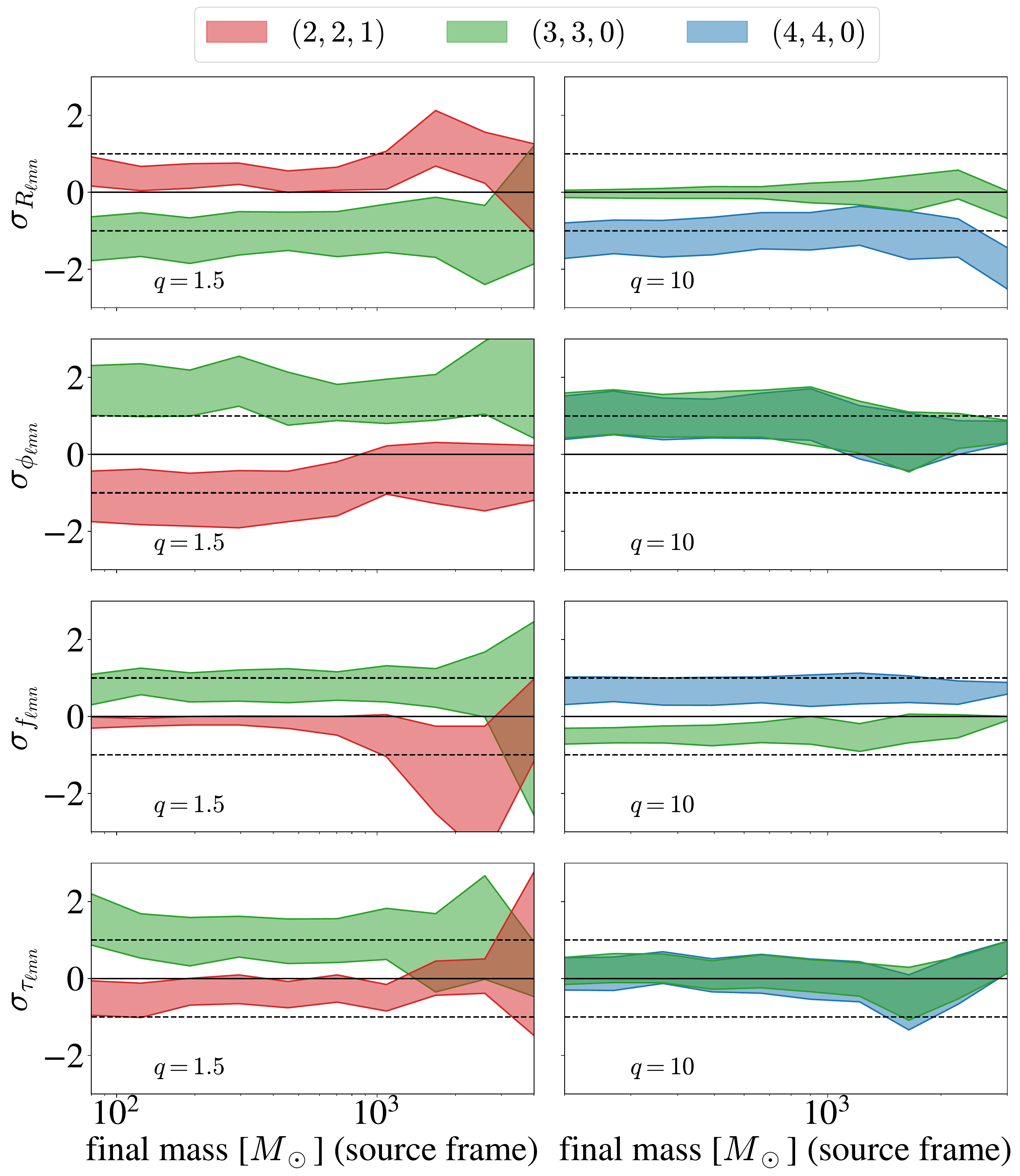}
    \caption{\emph{Identifying the secondary and tertiary modes:} same as Figure \ref{fig:errors_pars_2_modes}, but for the secondary and tertiary modes identified for the three-mode model $\mathcal{M}_3$: $(2,2,1)$ and $(3,3,0)$ for $q = 1.5$ and $(3,3,0)$ and $(4,4,0)$ for $q = 10$.}
    \label{fig:errors_pars_3_modes}
\end{figure}

In Figure~\ref{fig:errors_pars_3_modes} we show a similar analysis for the three-mode horizon, but only for the modes we identified as the secondary and tertiary modes in the three-mode model $\mathcal{M}_3$. For $q = 1.5$, the modes are identified as the $(2,2,1)$ and the $(3,3,0)$, whereas for $q = 10$ we have the $(3,3,0)$ and the $(4,4,0)$. The larger offsets that are more clearly visible in the deviations of some parameters of the tertiary mode are a result of the ``contamination'' from the next subdominant modes in the signal.

\section{Discussion: Rayleigh {\it versus} Bayes}
\label{sec:discussion}

Comparing Figures~\ref{fig:horizon_rayleigh_LIGO} and \ref{fig:horizon_bayes_2modes}, we can see that the BH spectroscopy horizons obtained with the Rayleigh criterion and with the Bayes factor threshold have similar trends. However, $z^{\rm spec, R}_{\ell mn}$ is consistently smaller than $z^{\rm spec, B}_{\ell mn}$. This can be due to two reasons: the Fisher matrix approximation is not valid at these large distances and/or the Rayleigh criterion is inherently too restrictive.

\begin{figure*}[!htb]
	\centering
	\includegraphics[width = 1.0\linewidth]{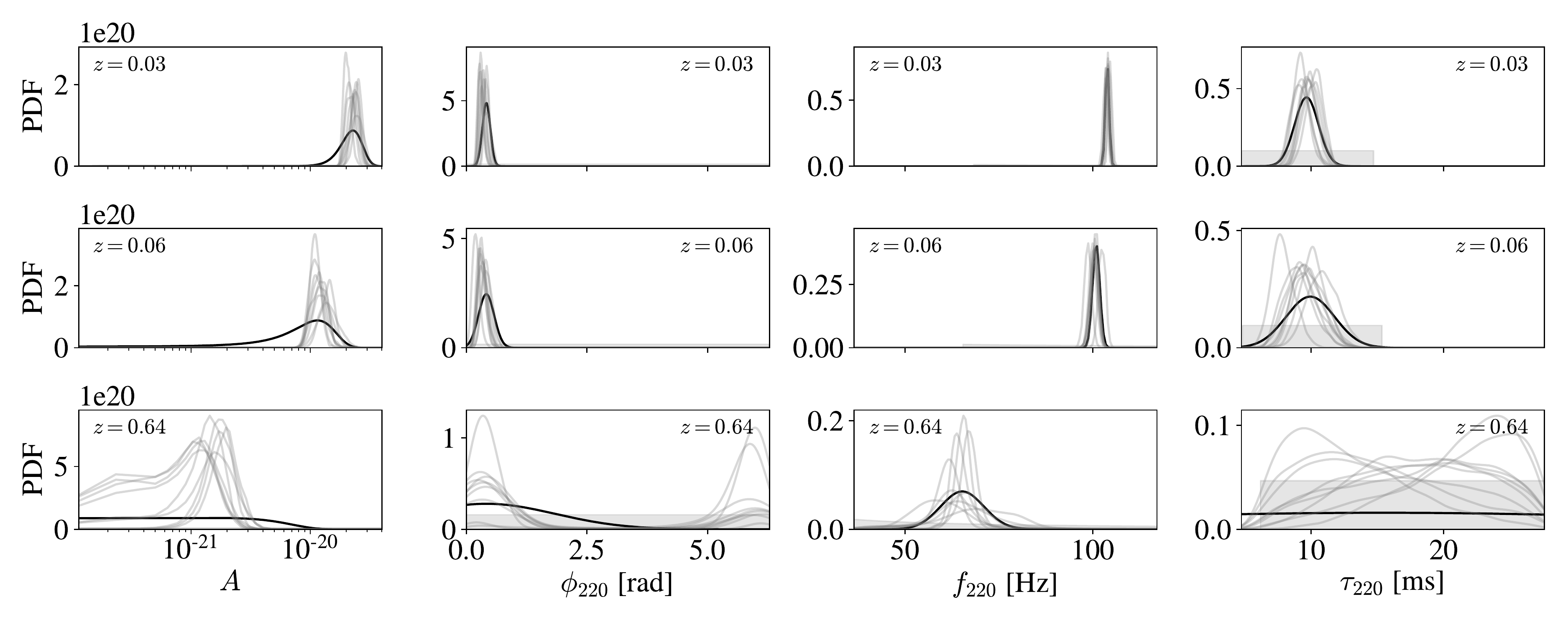}
	\includegraphics[width = 1.0\linewidth]{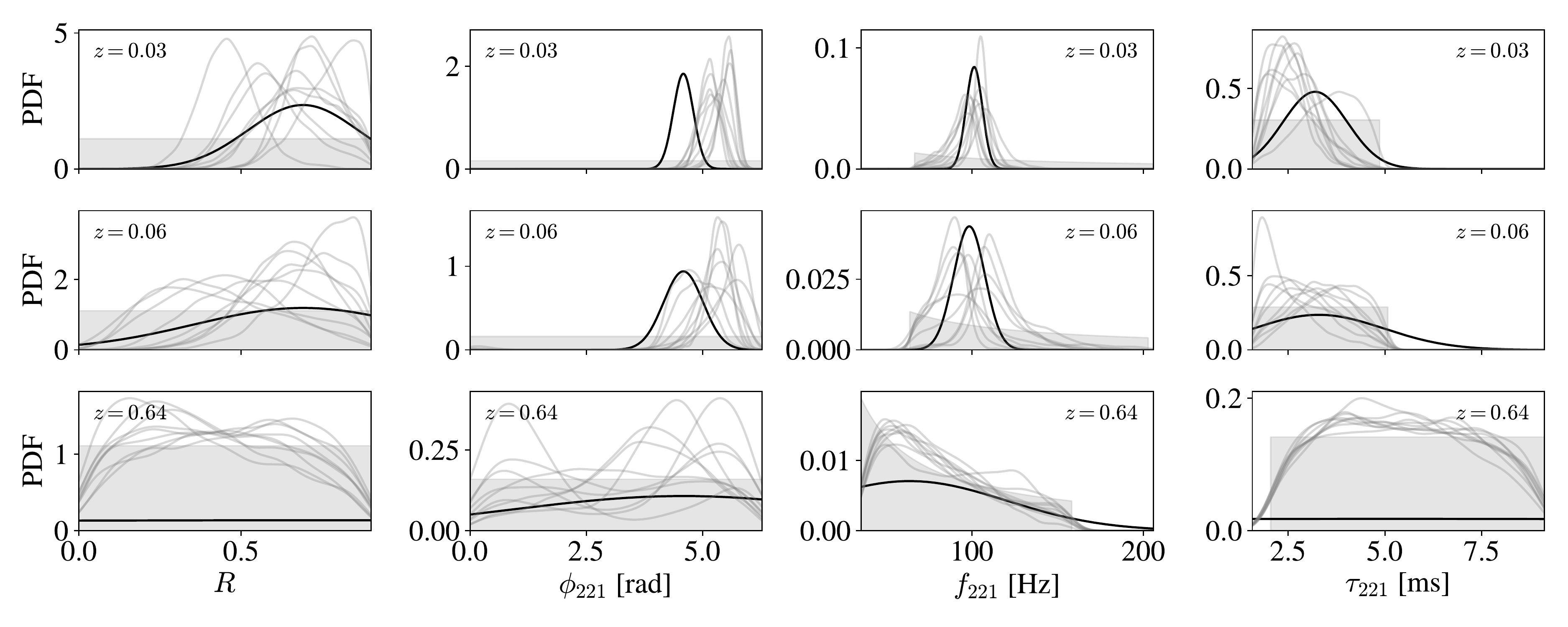}
	\caption{Posterior distributions for the eight parameters of a ringdown signal with the modes $(2,2,0)$ (top half of the figure) and $(2,2,1)$ (bottom half of the figure). We fixed the final BH mass $M = 156.3$ (similar to GW190521~\cite{Abbott:2020tfl}) and $q = 1.5$, and chose three distances: a distance inside the horizon $z^{\rm spec, B}_{221}$ (1st and 4th rows), the distance of the horizon $z^{\rm spec, B}_{221}$ (2nd and 5th rows) and a distance compatible with the event GW190521, well outside the horizon $z^{\rm spec, B}_{221}$ (3rd and 6th rows). The gray filled regions are the prior probability distributions used in the Bayes inference, the gray curves show 10 examples of the posterior probability distributions (obtained with independent noise realizations) for each case and the black curves represent a Gaussian distribution constructed with errors estimated by the Fisher matrix. The injected parameters correspond to the peak of the black curves.}
	\label{fig:fisher_parameters_gw190521}
\end{figure*}

The Fisher matrix analysis is only valid for high SNR signals \cite{Vallisneri:2007ev}, therefore we must be careful when using it for obtaining the BH spectroscopy horizons. Figure~\ref{fig:fisher_parameters_gw190521} shows the posterior distributions for the eight parameters of a ringdown signal with modes $(2,2,0)$ and $(2,2,1)$. We place the source at three distances: inside, outside and on the horizon $z^{\rm spec, B}_{221}$, in order to compare the example posteriors (grey curves) and the Fisher error estimation (black Gaussian) in each case. 
The agreement between the grey curves (within the variation provided by the noise) shows that the parameter estimation has converged. Additionally, the agreement of the grey curves with the black curve (which peaks at the injected value) shows that the injected parameters are recovered.

For the case \emph{inside} the horizon, the Fisher matrix error estimations agree well with the posteriors, but the errors are already noticeably larger for some parameters, such as $A$, $R$ and $\tau_{221}$. For the case \emph{on} the horizon, the Fisher matrix error for $\tau_{220}$ also becomes larger than the Bayes inference estimation. For the case \emph{outside} the horizon all of  the errors given by the Fisher Matrix are larger than the Bayes estimations, and for most parameters the black Gaussians appear nearly flat in the ranges shown. However, the posteriors for the $(2,2,1)$ parameters outside the horizon simply recover their priors, as the signal is not strong enough to be informative.

As expected, for large distances (low SNR) the Fisher matrix estimations are incorrect. For the case inside the Bayes factor horizon with $z=0.03$ the Fisher estimations are good, but this redshift is outside the Rayleigh criterion horizon $z^{\rm spec, R}_{221}$ for $M = 156.3$ (see Figure~\ref{fig:horizon_rayleigh_LIGO}). This indicates that the poor error predictions of the Fisher matrix are not solely responsible for restricting the horizons, and the Rayleigh criterion itself seems to be too restrictive to determine the evidence of a secondary mode. 

The very high SNR needed to satisfy both Rayleigh conditions was addressed by~\cite{Berti:2007zu}, where they proposed a generalized likelihood ratio test (GLRT) to identify the presence of a secondary mode in the signal. The GLRT is similar to the Bayes factor criterion, and in both criteria the confidence level can be increased in order to make them equivalent to the Rayleigh criterion, which results in a more restrictive detectability condition.

We computed the Bayes factor $\mathcal{B}^2_1$ for the $(2,2,1)$ mode at the corresponding Rayleigh horizon for the LIGO detector and $q = 1.5$, considering 500 noise realizations for 50 log-distributed masses. We found very high values, $\ln\mathcal{B}\gtrsim 60$ (or $\mathcal{B}\gtrsim10^{26}$). Moreover, the Bayes factor value is not constant at the Rayleigh horizon, it has a minimum near $M = 45$ and monotonically increases for smaller and larger masses. For higher harmonics, the Rayleigh criterion is more restrictive due to the damping time condition, resulting in even larger Bayes factors.

An alternative could be to require the resolvability of the quality factor $Q_{\ell m n} = \pi f_{\ell m n} \tau_{\ell m n}$ instead of the damping time $\tau_{\ell m n}$, or to require the resolvability of just one of the Rayleigh conditions, as done by~\cite{Bhagwat:2019dtm,Forteza:2020hbw}. 
In appendix~\ref{sec:appendix-other-criteria} we explore these and other variations of the Rayleigh criterion and compare them with the result of our Bayesian analysis for LIGO. 
A modification in the Rayleigh criterion that is compatible with the Bayes inference analysis is desirable for the extension of this kind of analysis for a larger set of systems, as the Bayesian analysis requires extensive computational resources.

\section{Conclusions}
\label{sec:conclusions}
The possibility of testing the no-hair theorem and GR with black hole spectroscopy is at hand. There is already some evidence for the detection of a secondary mode ~\cite{Isi:2019aib,Abbott:2020jks,Capano:2020dix}, and the detection of another event like GW150914 with current detector sensitivity could provide significantly improved constraints. We propose here a conservative criterion for detectability: a Bayes factor threshold $\ln \mathcal{B}^2_1 > 8$, which simultaneously guarantees that the secondary mode will be detectable in the signal and distinguishable from the fundamental mode $(2,2,0)$.

With this criterion we calculate the BH spectroscopy horizons, that is the maximum distance (averaged over sky location and binary inclination) up to which two or more modes can be detected in the ringdown of a binary black hole merger. Our analysis is restricted to non-spinning circular binaries with mass ratios equal to 1.5 and 10; the first mass ratio is compatible with most LVC detections~\cite{LIGOScientific:2018mvr,LIGOScientific:2020niy} and the second has an asymmetry that is important for the excitation of the higher harmonics in the ringdown. Importantly, we start analyzing all ringdown signals at $t = 10(M_1 + M_2)$ after the peak amplitude.

The best prospects for detection of a secondary mode arrive for nearly equal mass binaries and higher masses. In such cases, the overtone mode $(2,2,1)$ will be the easiest subdominant mode to detect, followed by the $(3,3,0)$ mode, as we determine in an agnostic multimode analysis. For very asymmetric binaries with a 10:1 mass ratio, the secondary and tertiary modes are the harmonics $(3,3,0)$ and $(4,4,0)$. At the high mass end of our analysis ($M \gtrsim 10^3\ M_{\odot}$), the secondary and tertiary modes are switched, as the modes with higher frequency stand out more against the low-frequency noise of the detector.

For LIGO, the expected rate for an event similar to GW190521 within the corresponding BH spectroscopy horizon redshift $z^{\rm spec, B}_{221} \sim 0.06$ is $0.03 - 0.10\ {\rm yr}^{-1}$. For CE, this rate is dramatically improved: $(0.6 - 2.4) \times 10^3\ {\rm yr}^{-1}$. Intermediate mass black holes with masses between approximately $10^2$ and $10^3$ solar masses have the largest horizons for ground-based detectors, but their rates are still unknown. Binary BH merger rates for LISA sources are also still very uncertain. However, LISA should have no problem resolving more than one extra mode in the ringdown.

Other proposals for appraising the resolvability of a secondary mode have been suggested in the literature. We present a detailed analysis of the Rayleigh criterion, and contrast its predictions with our Bayesian analysis. We find the Rayleigh criterion to be excessively restrictive, but useful for quicker estimates and establishing general trends. A correction of the Rayleigh criterion that could reproduce the Bayesian results would have many practical advantages. Different combinations and variations of the Rayleigh conditions provide constraints at varying levels of required significance (evidence) for the secondary mode.

Finally, a generalization of this work could include a network of detectors in the analysis, which would  increase the BH spectroscopy horizons by a factor $\lesssim 2$. Coherent mode stacking~\cite{Yang:2017zxs} has been proposed to use multiple measurements to increase the SNR of the subdominant quasinormal modes. The original proposal is not easily applied to overtones, but data from individual signals can be combined in a Bayesian framework \cite{2019PhRvD..99l4044Z} to determine how many ringdown events, with the current merger rate estimates, would be needed to detect the higher harmonics with an statistical evidence of $\ln \mathcal{B} > 8$.

\begin{acknowledgments}
We are thankful to Quentin Baghi for providing the LISA noise spectral density curve, Tito Dal Canton for answering our questions about LIGO noise and Lucas Sanches for help with Slurm/HPC. We also thank Cole Miller and the Flatiron CCA GW Group for useful discussions, and Emanuele Berti, Vitor Cardoso, Will Farr, Max Isi and Xisco Jimenez Forteza for comments on our manuscript.
IO was supported by grant 2018/21286-3 of the S\~ao Paulo Research Foundation (FAPESP). CC acknowledges support by NASA under award number 80GSFC17M0002. This work was completed at the Aspen Center for Physics, which is supported by National Science Foundation grant PHY-1607611. Resources supporting this work were provided by the NASA High-End Computing (HEC) Program through the NASA Center for Climate Simulation (NCCS) at Goddard Space Flight Center.
\end{acknowledgments}

\bibliography{main}

\appendix
\section{Variations of the resolvability criteria}
\label{sec:appendix-other-criteria}
Here we explore different variations of the Rayleigh criterion proposed in the literature and present the resulting BH spectroscopy horizons for each case.

The formulation of the Rayleigh conditions ~\eqref{eq:rayleigh_both} was first introduced in~\cite{Berti:2005ys}, where it was used to compute the ``critical'' SNR $\rho_{\rm crit}$ to resolve \emph{either} frequency or damping time and the larger SNR $\rho_{\rm both}$ needed to resolve \emph{both} conditions. It was found in~\cite{Berti:2005ys} that $\rho_{\rm crit}$ is one or two orders of magnitude smaller than $\rho_{\rm both}$. 

\begin{figure}[!htb]
	\centering
	\includegraphics[width = 1.0\linewidth]{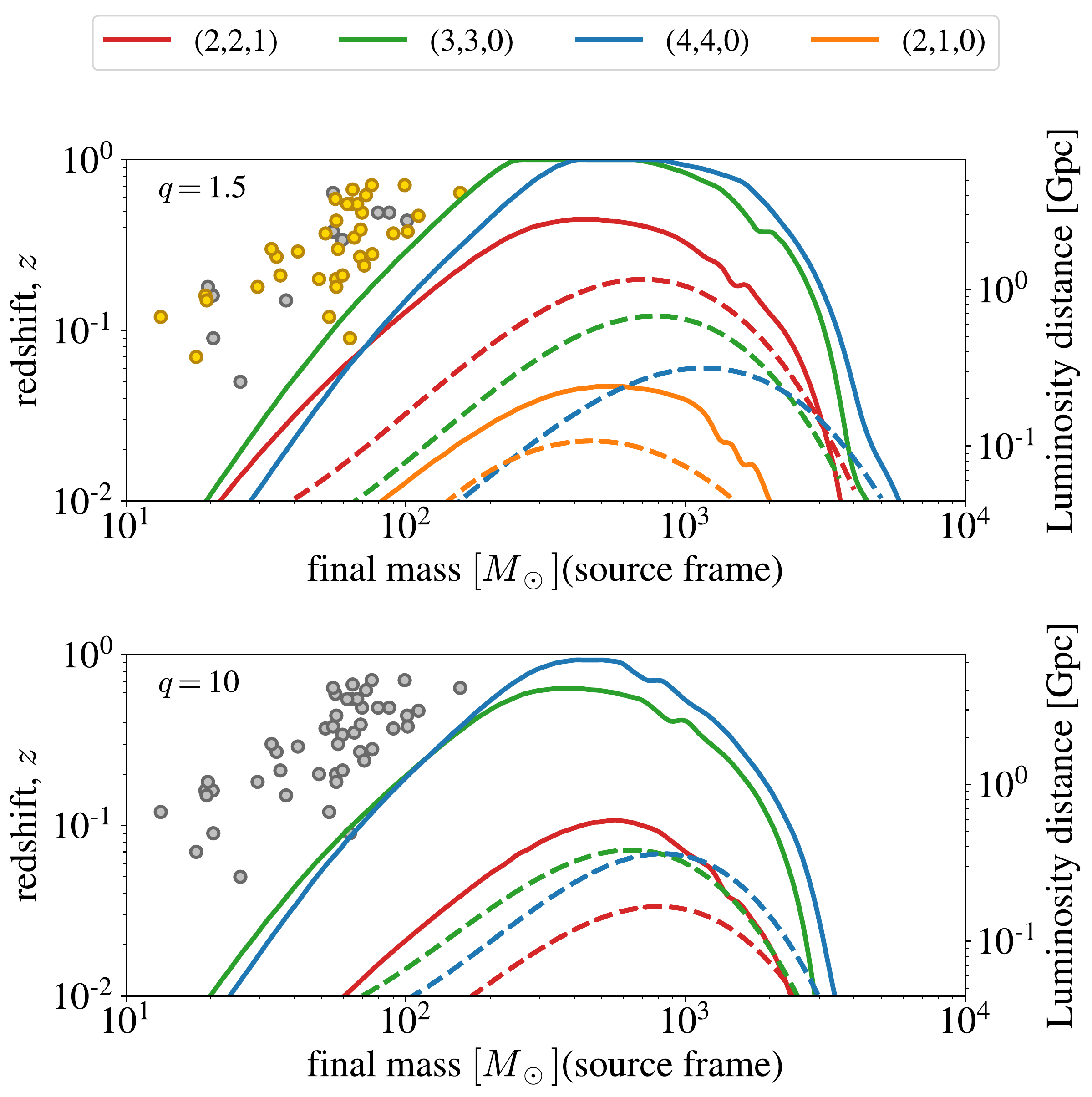}
	\caption{LIGO BH spectroscopy horizons obtained by requiring only one of Rayleigh conditions~\eqref{eq:rayleigh_f} or~\eqref{eq:rayleigh_tau} to be satisfied (solid lines). The errors are estimated with a Fisher matrix analysis. The dashed curves show the LIGO BH spectroscopy horizons obtained with a Bayes factor threshold $\ln \mathcal{B} > 8$ (same as in Figure~\ref{fig:horizon_bayes_2modes}). Circles show detections from GWTC-1 and GWTC-2; yellow circles are compatible with the mass ratio in each case, whereas gray circles are not. The one-condition Rayleigh horizons are much less restrictive and favor subdominant modes with $\ell \neq 2$.}
	\label{fig:appendix_rayleigh_one}
\end{figure}

We require that both conditions be satisfied in Section~\ref{sec:Rayleigh}. Some recent works have confirmed that requiring a single condition to be satisfied is not very restrictive~\cite{Bhagwat:2016ntk,Forteza:2020hbw}. Accordingly, we present in Figure~\ref{fig:appendix_rayleigh_one} the LIGO spectroscopy horizons by requiring the resolvability of the frequency \emph{or} the damping time. As expected, the horizons are much larger than in the case when both conditions are required, shown in Figure~\ref{fig:horizon_rayleigh_LIGO}. We can see that requiring just one condition favors the modes with $\ell \neq 2$, which have easily resolvable frequencies. The $(2,2,1)$ horizon is further reduced by the very large errors in the damping times estimated by the Fisher matrix analysis at the horizon distances (see discussion in Section \ref{sec:discussion}).

\begin{figure}[!htb]
	\centering
	\includegraphics[width = 1.0\linewidth]{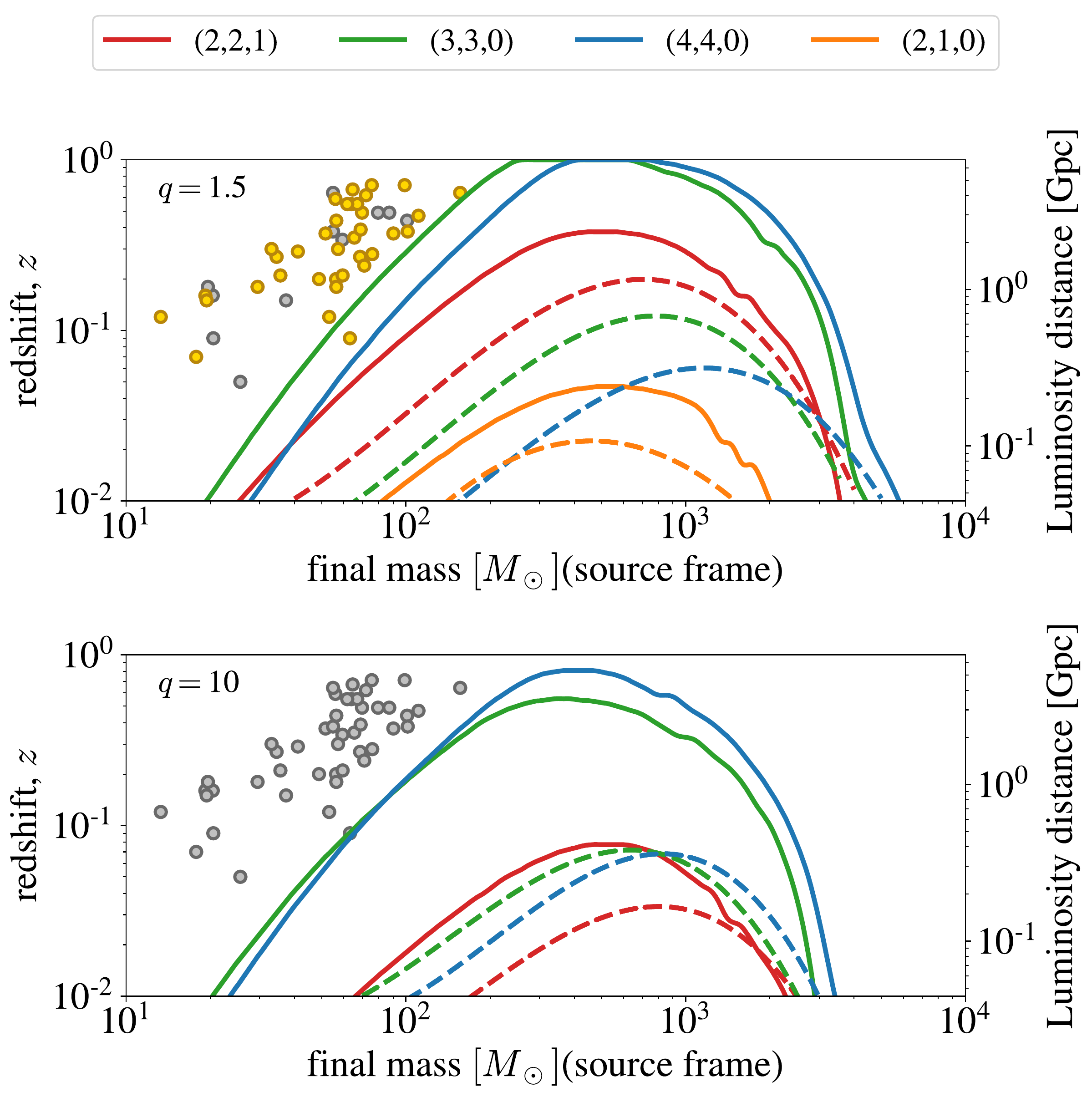}
	\caption{Same as Figure \ref{fig:appendix_rayleigh_one}, but here the solid lines show the LIGO BH spectroscopy horizons obtained with condition~\eqref{eq:appendix_resolvability_posterior}. These horizons are very similar to the single-condition Rayleigh horizons of Figure~\ref{fig:appendix_rayleigh_one}.}
	\label{fig:appendix_resolvability_posterior}
\end{figure}

More recently, \cite{2021arXiv210705609I} reinforced the idea that requiring both conditions should not be necessary to distinguish between two ringdown modes and suggested a single condition for the distinguishability of the 2-dimensional posteriors of frequencies and damping times,
\begin{equation}
	\frac{(f_{\ell m n} - f_{\ell' m'n'})^2}{\sigma^2_{f_{\ell m n}} + \sigma^2_{f_{\ell' m' n'}}} + \frac{(\tau_{\ell m n} - \tau_{\ell' m'n'})^2}{\sigma^2_{\tau_{\ell m n}} + \sigma^2_{\tau_{\ell' m' n'}}} \gtrsim 1.
	\label{eq:appendix_resolvability_posterior}
\end{equation}
In Figure~\ref{fig:appendix_resolvability_posterior} we show the LIGO BH spectroscopy horizons obtained with equation~\eqref{eq:appendix_resolvability_posterior}. We can see that these horizons are very close to the single-condition Rayleigh horizons in Figure~\ref{fig:appendix_rayleigh_one} and the same considerations apply here. If the Fisher matrix error estimates are replaced with values derived from the posterior distributions, the $(2,2,1)$ horizon reaches farther and the constraints become compatible with a somewhat lower Bayes factor threshold than the conservative value we used. Obviously, in each \emph{individual} event detected, properties such as the mass ratio, sky location and binary inclination will also impact the detectability of a secondary mode.

In~\cite{Forteza:2020hbw} the use of the quality factor $Q_{\ell m n } = \pi f_{\ell m n }\tau_{\ell m n }$ instead of the damping time was proposed, replacing equation~\eqref{eq:rayleigh_tau} by
\begin{equation}
	|Q_{\ell mn} - Q_{\ell' m'n'}|  > {\rm max}(\sigma_{Q_{\ell mn}},\sigma_{Q_{\ell' m'n'}}).
	\label{eq:appendix_Q_rayleigh}
\end{equation}
The damping times are also replaced by the quality factors for the Fisher matrix parameters, which results in different error estimation for the frequencies.

\begin{figure}[!htb]
	\centering
	\includegraphics[width = 1.0\linewidth]{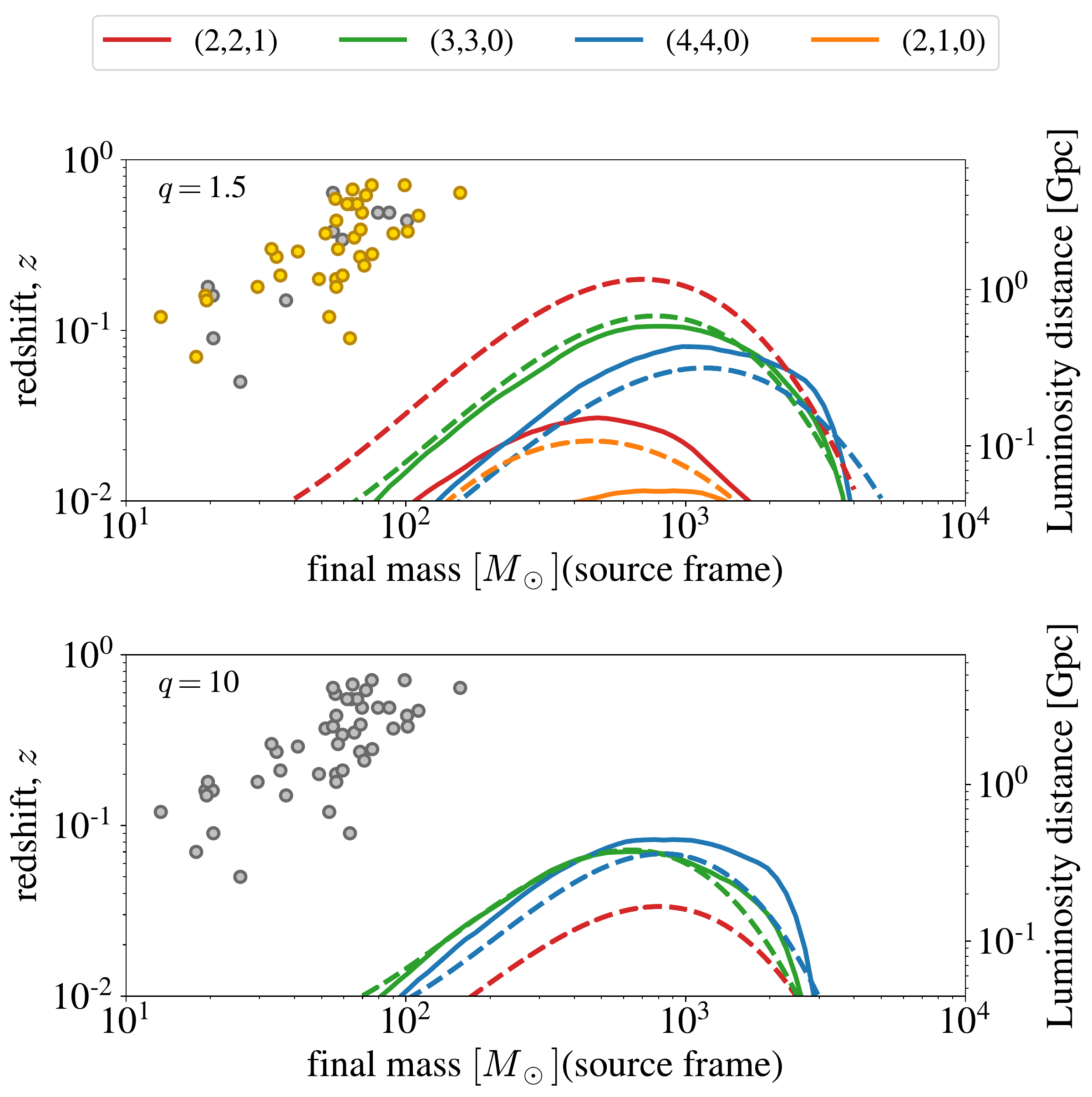}
	\caption{Same as Figure \ref{fig:appendix_rayleigh_one}, but using the quality factors $Q_{\ell m n } = \pi f_{\ell m n }\tau_{\ell m n }$ instead of the damping times in the Rayleigh conditions~~\eqref{eq:rayleigh_both} (solid lines). The horizons obtained for modes with $\ell \neq 2$ are very close to the horizons obtained with the Bayes factor threshold (dashed lines), but the horizons of the modes with $\ell = 2$ are more restrictive.}
	\label{fig:appendix_Q_rayleigh}
\end{figure}

In Figure~\ref{fig:appendix_Q_rayleigh} we show the LIGO spectroscopy horizons by requiring the resolvability of \emph{both} frequencies and quality factors. The horizons for the harmonics $(3,3,0)$ and $(4,4,0)$ are very close to the Bayes factor horizons, but the horizons of the overtone and the $(2,1,0)$ mode are still very restrictive. This is expected, as the quality factor ``relaxes'' the second Rayleigh condition on the fundamental higher harmonic modes, which have damping times very similar to the dominant mode $(2,2,0)$.

In Figures \ref{fig:appendix_rayleigh_one}, \ref{fig:appendix_resolvability_posterior} and \ref{fig:appendix_Q_rayleigh} we impose different prescriptions for the resolvability criteria of two modes in the ringdown, in the spirit of the original Rayleigh criterion. However, we do not impose the  detectability requirement that each mode $(\ell, m, n)$ should have a SNR $\rho_{\ell m n} > 8$, which we also impose in our Rayleigh horizons in Section~\ref{sec:Rayleigh}. (In Section~\ref{sec:Rayleigh}, however, this condition is superfluous: requiring both Rayleigh conditions to be satisfied is more restrictive than the SNR requirement.)

\begin{figure}[!htb]
	\centering
	\includegraphics[width = 1.0\linewidth]{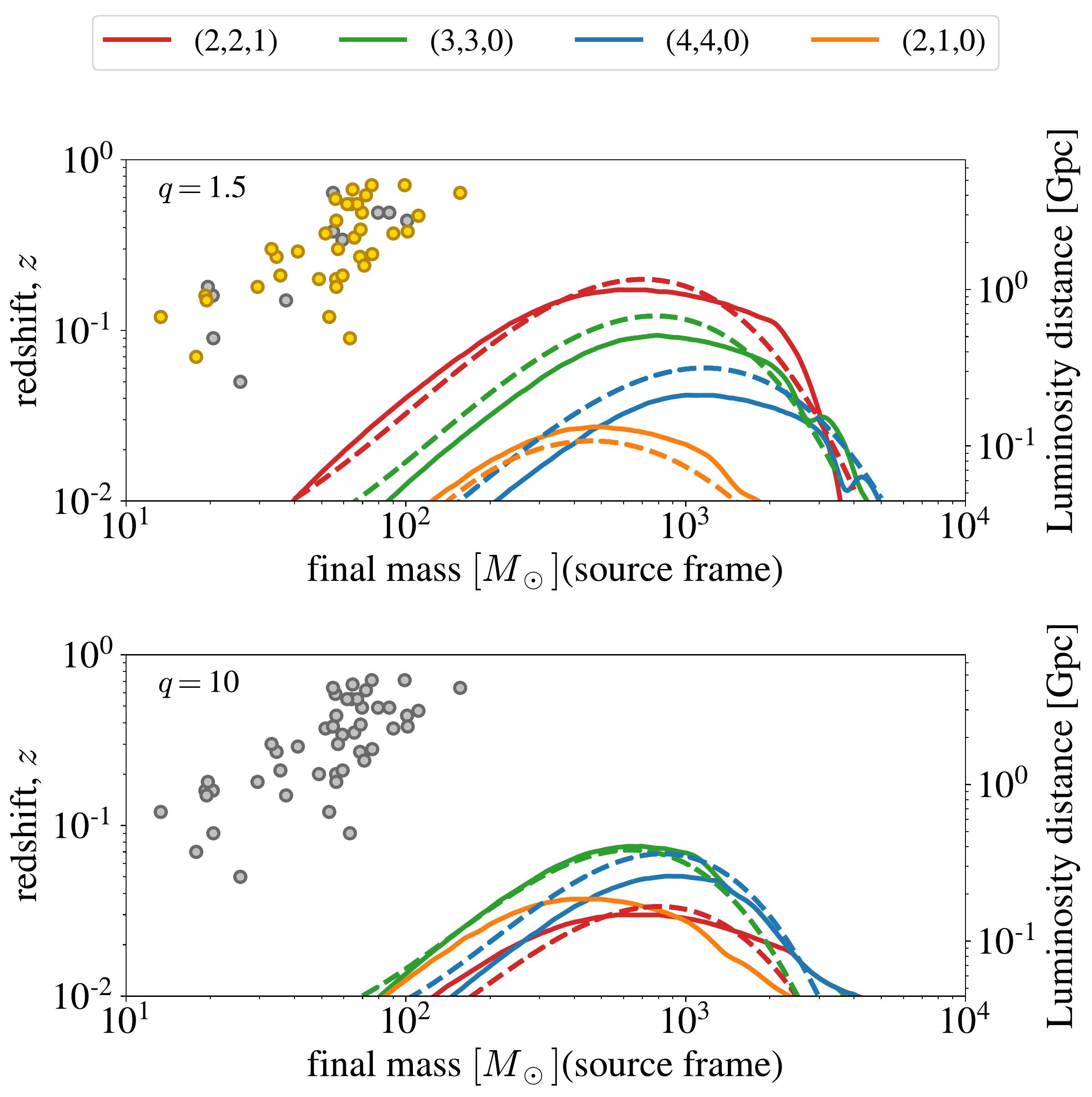}
	\caption{Same as Figure \ref{fig:appendix_rayleigh_one}, but this time we show with solid lines the LIGO SNR horizons obtained by requiring $\rho_{\ell m n} > 8$ for each subdominant mode $(\ell, m,n)$. The SNR horizons are close to the Bayes factor threshold, with the exception of the $(2,1,0)$ mode for $q=10$.}
	\label{fig:appendix_snr}
\end{figure}

In Figure~\ref{fig:appendix_snr} we show the SNR = 8 horizons, which are much more restricted than the horizons in Figures~\ref{fig:appendix_rayleigh_one} and~\ref{fig:appendix_resolvability_posterior}. The SNR criterion estimates horizons very close to the Bayes factor horizon, indicating that a SNR threshold for the subdominant modes is necessary for a high Bayes factor, but it is not sufficient, as we can see that the $(2,1,0)$ SNR horizon for $q = 10$ is very large. We note that our SNR horizons are somewhat smaller than the corresponding ringdown horizons reported for fundamental modes by \cite{Baibhav:2018rfk} due to the different mode amplitudes assumed in our analysis, as we take the start time to be $t = t_{\rm peak} + 10 M$ (see also~\cite{Baibhav:2017jhs}).

Finally, there are additional resolvability criteria proposed in the literature that we do not explore here. In~\cite{2021arXiv210705609I}, it is proposed that a subdominant mode is detected if its amplitude posterior excludes zero in the 90\% credible interval, and they find that this condition is satisfied for the first overtone of a GW150914-like event; this was also confirmed by other works~\cite{Bustillo:2020tjf,Isi:2019aib,Isi:2020tac}. In~\cite{Forteza:2020hbw}, the authors require the similar condition $\sigma_R < R$, for the mode amplitude ratio $R$ to exclude zero at the 1$\sigma$ level, and they also propose a ``measurability'' criterion, which requires a relative accuracy for the estimated QNMs parameters.
\clearpage
\end{document}